\documentclass[aps,showpacs,nofootinbib,superscriptaddress]{revtex4}
\usepackage{graphicx}
\usepackage{dcolumn}
\usepackage{multirow}

 \def\tstrut{\vrule height2.5ex depth0pt width0pt} 

\begin{document}

\title{ Static properties and semileptonic decays of doubly heavy
baryons in a nonrelativistic quark model. } \author{C.
Albertus}\thanks{Present address: School of Physics and Astronomy, University of
Southampton, Southampton SO17 1BJ, United Kingdom.}
\affiliation{Departamento de F\'{\i}sica At\'omica, Molecular y
Nuclear, Universidad de Granada, E-18071 Granada, Spain.}  \author{
E. Hern\'andez} \affiliation{Grupo de F\'\i sica Nuclear, Departamento
de F\'\i sica Fundamental e IUFFyM, Facultad de Ciencias, E-37008
Salamanca, Spain.}  \author{J. Nieves} \affiliation{Departamento de
F\'{\i}sica At\'omica, Molecular y Nuclear, Universidad de Granada,
E-18071 Granada, Spain.} \author{ J. M. Verde-Velasco}
\affiliation{Grupo de F\'\i sica Nuclear, Departamento de F\'\i sica
Fundamental e IUFFyM, Facultad de Ciencias, E-37008 Salamanca, Spain.}
\begin{abstract} 
\rule{0ex}{3ex} We evaluate static properties and semileptonic decays
for the ground state of doubly heavy $\Xi,\,\Xi',\,\Xi^*$ and 
$\Omega,\,\Omega',\,\Omega^*$ baryons. Working in the framework of a nonrelativistic quark
model, we solve the three--body problem by means of a variational
ansatz made possible by heavy quark spin symmetry constraints. To
check the dependence of our results on the inter-quark interaction we
use five different quark-quark potentials that include a confining
term plus Coulomb and hyperfine terms coming from one--gluon
exchange. Our results for static properties (masses, charge 
radii and magnetic moments ) are, with a few exceptions for the
magnetic moments, in good agreement with a previous Faddeev
calculation. Our much simpler wave functions are  used to evaluate
semileptonic decays of doubly heavy $\Xi,\Xi'(J=1/2)$ and $\Omega,
\Omega'(J=1/2)$ baryons. Our results for the decay widths are in good
agreement with calculations done within a relativistic quark model in
the quark--diquark approximation.
\end{abstract}

\maketitle

\section{Introduction}

%
\begin{table}

\begin{tabular}{cccccccc}\hline
Baryon &~~~~$S$~~~~&~~~~$J^P$~~~~&~~~~ $I$~~~~&~~~~$S_{ h}^\pi$~~~~& 
Quark content 

\\
       &       &         &   &          &                 
\\\hline
$\Xi_{cc}$& 0 &$\frac12^+$& $\frac12$ &$1^+$&$ccl$
\\
$\Xi^*_{cc}$ & 0 &$\frac32^+$&$\frac12$  &$1^+$&$ccl$
\\
$\Omega_{cc}$ & $-1$ &$\frac12^+$& 0 &$1^+$&$ccs$
\\
$\Omega^*_{cc}$ & $-1$ &$\frac32^+$&$0$&$1^+$&$ccs$
\\\\
$\Xi_{bb}$& 0 &$\frac12^+$& $\frac12$ &$1^+$&$bbl$
\\
$\Xi^*_{bb}$ & 0 &$\frac32^+$&$\frac12$  &$1^+$&$bbl$
\\
$\Omega_{bb}$ & $-1$ &$\frac12^+$& 0 &$1^+$&$bbs$
\\
$\Omega^*_{bb}$ & $-1$ &$\frac32^+$&$0$&$1^+$&$bbs$
\\\\
$\Xi'_{bc}$& 0 &$\frac12^+$& $\frac12$ &$0^+$&$bcl$
\\
$\Xi_{bc}$ & 0 &$\frac12^+$&$\frac12$  &$1^+$&$bcl$
\\
$\Xi^*_{bc}$& 0 &$\frac32^+$& $\frac12$ &$1^+$&$bcl$
\\
$\Omega'_{bc}$ & $-1$ &$\frac12^+$& 0 &$0^+$&$bcs$
\\
$\Omega_{bc}$ & $-1$ &$\frac12^+$& 0 &$1^+$&$bcs$
\\
$\Omega^*_{bc}$ & $-1$ &$\frac32^+$&$0$&$1^+$&$bcs$
\\
\hline
\end{tabular}
\caption{Quantum numbers of doubly heavy baryons analyzed in this study. $S$, $J^P$ are strangeness and the spin parity
of the baryon, $I$ is the isospin, and
$S_{h}^\pi$ is the spin parity of the heavy
degrees of freedom. $l$ denotes a light  $u$ or
$d$ quark .}
\label{tab:summ}
\end{table}



The subject of doubly heavy baryons has been
attracting attention for a long time. Magnetic moments of doubly
charmed baryons were evaluated back in the 70's by
Lichtenberg~\cite{lichtenberg77} within a nonrelativistic approach.
The infinite heavy quark mass limit was already used in the 90's to
relate the spectrum of doubly heavy baryons to the one of mesons with
a single heavy quark~\cite{savage90}, or to analyze their semileptonic
decay~\cite{white91}. A factor of two error in the hyperfine splittings of Ref.~\cite{savage90} 
 have been recently noticed 
by the potential nonrelativistic QCD (pNRQCD) calculation of Ref.~\cite{brambilla05}.

On the experimental side the SELEX Collaboration claimed
evidence for  the $\Xi^+_{cc}$ baryon, in the $\Lambda_c^+K^-\pi^+$ and 
$pD^+K^-$ decay modes, with a mass of
$M_{\Xi^+_{cc}}=3519\pm 1\ \mathrm{MeV/c^2}$~\cite{mattson02}. Those results
were challenged by a theoretical analysis~\cite{kili} which claimed 
 the observed events
by the SELEX Collaboration could be explained without the involvement
of doubly charmed baryons. Other experimental collaborations like 
FOCUS~\cite{focus03}, {\sl BABAR}  ~\cite{babar06} and BELLE~\cite{lesiak06}
have found no evidence for doubly charmed baryons so far.   At present
 the    $\Xi^+_{cc}$ has only a one star status and it  is not listed in  the
particle summary table~\cite{pdg06}.

In hadrons with a heavy quark and working in the
infinite heavy quark mass limit the dynamics of the light degrees of
freedom becomes independent of the heavy quark flavor and spin. This
is known as heavy quark symmetry (HQS)~\cite{hqs1,hqs2,hqs3,hqs4}.
This symmetry was developed into an effective theory
(HQET)~\cite{georgi90} that allowed a systematic, order by order,
evaluation of corrections in inverse powers of the heavy quark masses.
Unfortunately ordinary HQS can not be applied directly to hadrons
containing two heavy quarks as the kinetic energy term needed in those
systems to regulate infrared divergences breaks heavy flavor
symmetry~\cite{thacker91}.  For those hadrons the symmetry that
survives is heavy quark spin symmetry (HQSS)~\cite{jenkins93}, which
amounts to the decoupling of the heavy quark spins in the infinite
heavy quark mass limit. In that limit one can consider the total spin
of the two heavy quark  subsystem ($S_h$) to be well defined.  In this
work we shall assume this is a good approximation for the actual heavy
quark masses. This approximation, which is the only one related to the
infinite heavy quark mass limit that we shall use, will certainly
simplify the solution of the baryon three--quark problem. Recently the authors of
Ref.~\cite{brambilla05} have developed and effective theory (pNRQCD) suitable to
describe baryons with two and three heavy quarks.

Solving the three--body problem is not an easy task and here we shall
do it by means of a variational approach. The approach, with obvious
changes, was already applied with good results in the study of baryons
with one heavy quark~\cite{conrado04}. This method, that leads to
simple and manageable wave functions, is made possible by the
simplifications introduced in the problem by the fact that we can
consider $S_h$ to be well defined.  We shall consider several simple
phenomenological quark--quark interactions~\cite{BD81,SS94,Si96} which
free parameters have been adjusted in the meson sector and are thus
free of three--body ambiguities. The use of different interactions
will allow us to estimate part of the theoretical uncertainties
affecting   our calculation. 
Uncertainties related to the nonrelativistic 
 baryon states that we use are difficult to estimate. 
We are aware of the limitations of a non-relativistic approach to
   describe light quark physics. However, for the kind of study
   performed in this work these are likely not as relevant as in other
   contexts. We study the semileptonic decays of double heavy baryons,
   in which the light quark is merely an spectator (heavy-to-heavy
   transitions). Indeed the lack of a proper relativistic treatment
   did prevent us to study semileptonic transitions of the type $b \to
   u$ of a greater phenomenological interest. 
 One should notice
however that at least part of the relativistic effects not explicitly taken into 
account are included in an effective way in the parameters of the interaction
 which had been  fitted to
experimental data. We think this explains why the nonrelativistic quark model is
so successful  phenomenologically even in the presence of light quarks.

Our simple variational calculation reproduces the results for static
properties obtained in Ref.~\cite{Si96} by solving more involved
Faddeev type equations. Our method has the  advantage that we provide explicit and
manageable wave
functions that can be used to evaluate further observables. Static properties like masses and magnetic
moments of doubly heavy baryons have also been studied in other
models.  Masses have been calculated in the relativistic quark model
assuming a light quark heavy diquark structure~\cite{ebert02}, the
potential approach and sum rules of QCD~\cite{kiselev02}, the
nonperturbative QCD approach~\cite{narodetskii02}, the Bethe--Salpeter
equation applied to the light quark heavy diquark~\cite{tong00}, the
nonrelativistic quark model with harmonic oscillator
potential~\cite{itoh00} or with the use of QCD derived
potentials~\cite{vijande04,gershtein00}, the relativistic quasi--potential quark
model~\cite{ebert97}, with the use of the Feynman-Hellman theorem and
semi-empirical mass formulas within the framework of a nonrelativistic
constituent quark model~\cite{roncaglia95},  in
effective field theories~\cite{korner94,brambilla05}, or in lattice nonrelativistic
QCD~\cite{mathur02}. There are also lattice QCD determinations
~\cite{khan00,lewis01,flynn03}. Similarly, magnetic moments have
been evaluated in a nonrelativistic approach~\cite{lichtenberg77}, in
the relativistic three--quark model~\cite{faessler06}, the
relativistic quark model using different forms of the relativistic
kinematics~\cite{bruno04}, in the skyrmion model~\cite{oh91}, in the
Dirac equation formalism~\cite{jena86}, or using the MIT bag
model~\cite{bose80}.

We shall further use our manageable wave functions to study
semileptonic decays of doubly heavy , $J=1/2$, baryons. We shall evaluate
form factors, decay widths and angular asymmetry parameters. Previous
calculations of semileptonic decay widths have been done in different
relativistic quark model approaches ~\cite{ebert04,faessler01,guo98},
or with the use of HQET~\cite{sanchis95}.

The paper is organized as follows. In Sect.~\ref{sec:3bp} we study the
Hamiltonian of the system (Subsect.~\ref{subsec:h}) and briefly
introduce the different inter-quark interactions used in this work
(Subsect.~\ref{subsec:qqi}).  The variational wave functions are
discussed in Sect.~\ref{sec:vwf}. In Sect.~\ref{sec:sp} we present
results for the static properties: masses
(Subsect. \ref{subsec:masses}), charge 
densities and radii
(Subsect.~\ref{subsec:chmdr}), and magnetic moments
(Subsect.~\ref{subsec:mm}). Semileptonic decays are analyzed in
Sect.~\ref{sec:sd}. After the presentation of general formulas, in
Subsect.~\ref{subsec:ff} we relate the form factors to matrix elements
and show how the latter ones are evaluated within our model. In
Subsect.~\ref{subsec:results} we present our results for the form
factors, differential and total semileptonic decay widths, and angular
asymmetry parameters. The findings of this work are summarized in
Sect.~\ref{sec:summary}. The paper also includes three appendices: in
 appendix
~\ref{app:ihqml} we analyze the infinite heavy quark mass limit of our 
radial wave functions. In
appendix~\ref{app:fg123} we relate the form factors for semileptonic
decay to the two basic integrals in terms of which all of them can be
obtained. Finally, in appendix~\ref{app:integrals} we give explicit
expressions for those basic integrals.

  In Table~\ref{tab:summ} we summarize the quantum numbers of the
doubly heavy baryons considered in this study\footnote{Note that
 the definitions 
of $\Xi_{bc}$ and
$\Xi'_{bc}$ are interchanged in some references, with
$\Xi_{bc}$ having $S_h=0$ and $\Xi'_{bc}$ having $S_h=1$.  The same
applies to $\Omega_{bc}$ and $\Omega'_{bc}$.  In tables we always quote the
results corresponding to the convention we use.}.

\section{Three Body Problem}
\label{sec:3bp}
\subsection{Intrinsic Hamiltonian}
\label{subsec:h}
\begin{figure}[t]
\resizebox{10.cm}{7.cm}{\includegraphics{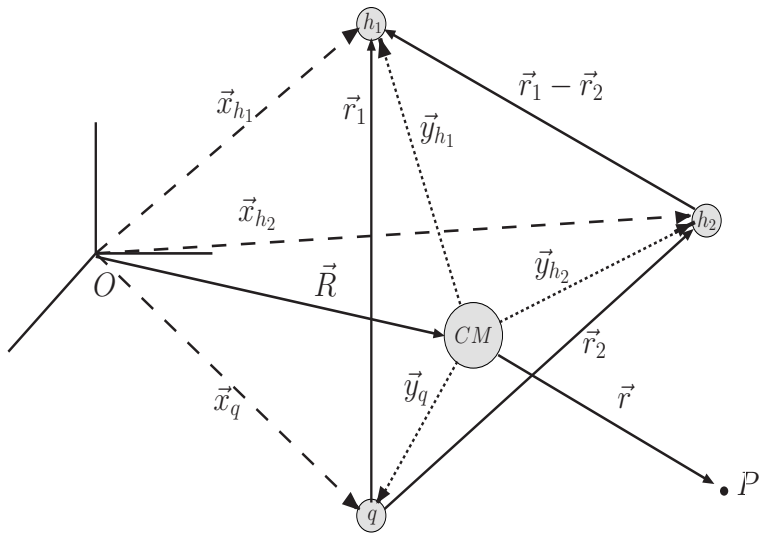}}
\caption{ Definition of different coordinates used throughout this
work.}\label{fig:coor}
\end{figure}

In the Laboratory (LAB) frame (see Fig.~\ref{fig:coor}), the
Hamiltonian ($H$) of the three quark ($h_1,\,h_2,\,q$, where
$h_1,\,h_2=c,b$ and $q=l(u,\,d),\,s$) system reads:
\begin{eqnarray}
H&=&  \sum_{j=h_1,\,h_2,\,q} \left (m_j
-\frac{\stackrel{\rightarrow}{\nabla}\stackrel{}{^2}_{\hspace{-.1cm}\vec x_j}}{2m_j}\right ) +
V_{h_1h_2} + V_{h_1q}+V_{h_2 q }
\end{eqnarray}
where $m_{h_1},\,m_{h_2},\,m_q$  are the quark masses and the
quark-quark interaction terms $V_{jk}$ depend on the quark
spin-flavor quantum numbers and the quark coordinates ($\vec{x}_{h_1},
\vec{x}_{h_2},\, \vec{x}_q$ for the $h_1,\,h_2,\,q$ quarks
respectively). To separate the Center of Mass (CM) free motion,
we go to the light quark frame ($\vec{R},\vec{r}_1,\vec{r}_2$),
\begin{eqnarray}
\vec{R}&=&\frac{m_{h_1}\vec{x}_{h_1} + m_{h_2}\vec{x}_{h_2} + m_q \vec{x}_q}
{m_{h_1}+m_{h_2}+m_q} \nonumber \\
\vec{r}_1 & = & \vec{x}_{h_1} - \vec{x}_q \nonumber \\
\vec{r}_2 & = & \vec{x}_{h_2} - \vec{x}_q 
\end{eqnarray}
where $\vec{R}$ and $\vec{r}_1,\, \vec{r}_2$ are the CM position in
the LAB frame and the relative positions of the $h_1,\,h_2$ heavy quarks 
with respect to the light quark $q$.   The Hamiltonian now reads
\begin{eqnarray}
H&=&
-\frac{\stackrel{\rightarrow}{\nabla}
\stackrel{}{^2}_{\hspace{-.1cm}\vec{R}}}{2 \overline M} +
H^{\rm int} 
\end{eqnarray}
\begin{eqnarray}
 H^{\rm
int}&=&\overline M+\sum_{j=1,2}\,H_j^{sp}+V_{h_1h_2}(\vec r_1-\vec r_2,\, spin)-
\frac{\stackrel{\rightarrow}{\nabla}_1\cdot\stackrel{\rightarrow}{\nabla}_2}
{m_q}
\nonumber\\
H_j^{sp}&=&-\frac{\stackrel{\rightarrow}{\nabla}\stackrel{}{^2}_{\hspace{-0.2cm}j}
}{2\mu_j}+V_{h_jq}(\vec r_j,\, spin),\ j=1,2
\label{eq:hint}
\end{eqnarray}
where $\overline M = m_{h_1}+m_{h_2}+m_{q}$, $\mu_{j} = \left (
1/m_{h_j} + 1/m_q\right)^{-1}$ and $\stackrel{\rightarrow}{\nabla}_{j}
= \partial/\partial_{\vec{r}_j},\ j=1,2$. The intrinsic Hamiltonian
$H^{\rm int}$ describes the dynamics of the baryon. Apart from the sum of the
 quark masses $\overline M$, it consists of the
sum of two single particle Hamiltonian $H^{sp}_j$, each of them
describing the dynamics of a heavy-light quark system, plus the
heavy--heavy interaction term, including the Hughes-Eckart term
($\stackrel{\rightarrow}{\nabla}_1\cdot
\stackrel{\rightarrow}{\nabla}_2$).  We will use a variational approach to solve it.

\subsection{Quark--Quark Interactions}
\label{subsec:qqi} 

We have examined five different interactions, one suggested by Bhaduri
and collaborators~\cite{BD81} (BD) and four suggested by
B. Silvestre-Brac and C. Semay~\cite{Si96,SS94} (AL1, AL2, AP1 y
AP2). All of them contain a confinement term, plus Coulomb and
hyperfine terms coming from one-gluon exchange, and differ from each
other in the form factors used for the hyperfine terms, the power of
the confining term or the use of a form factor in the one gluon
exchange Coulomb potential.  All free parameters in the potentials had
been adjusted to reproduce the light ($\pi$, $\rho$, $K$, $K^*$, etc.)
and heavy-light ($D$, $D^*$, $B$, $B^*$, etc.) meson
spectra\footnote{To get the quark--quark interaction starting from a
quark-antiquark one the usual $V_{ij}^{qq} = V_{ij}^{q \bar q}/2$
prescription, coming from a $\vec{\lambda}_i\vec{\lambda}_j$ color
dependence ($\vec{\lambda}$ are the Gell-Mann matrices) of the whole
potential, has been assumed.  }. All details on the above interactions
can be found in Refs.~\cite{BD81,Si96,SS94}.

These interactions were also used in Ref.~\cite{Si96} to obtain,
within a Faddeev calculation, the spectrum and static properties of
heavy baryons. Our simpler variational method will not only give
equally good results for the observables analyzed in~\cite{Si96}, but
it will also provide us with easy to handle wave functions that can be
used to evaluate other observables.

\section{Variational Wave Functions}\label{sec:vwf}
For the above mentioned interactions,  we have that both the total spin  and
 the internal orbital angular momentum given
 as
\begin{eqnarray}
{\vec
S} &=& \left ( {\vec \sigma}_{h_1} + {\vec \sigma}_{h_2} +
{\vec \sigma}_q \right )/2\nonumber\\
{\vec L} &=& {\vec l}_1 + {\vec l}_2, \qquad {\rm with}~~ {\vec
l}_j=-{\rm i}~~{\vec r}_j \times {\stackrel{\rightarrow}{\nabla}}_j, \quad j=1,2
\end{eqnarray}
commute with the intrinsic Hamiltonian and are thus well defined.  We
are interested in the ground state of baryons with total angular
momentum $J=1/2,\,3/2$\ so that we can assume the orbital angular
momentum of the baryons to be $L=0$.  This implies that the spatial
wave function can only depend on the relative distances $r_1$, $r_2$
and $r_{12}=|\vec{r}_1-\vec{r}_2|$. Furthermore when the heavy quark
mass is infinity ($m_h \to \infty$), the total spin of the heavy
degrees of freedom, $\vec{S}_{\rm heavy} = \left ( {\vec \sigma}_{h_1}
+ {\vec \sigma}_{h_2} \right )/2$, commutes with the intrinsic
Hamiltonian, since the spin--spin terms in the potentials vanish in
this limit. We can then assume the spin of the heavy degrees of
freedom to be well defined.

 With these simplifications we have used the following intrinsic wave
functions in our variational approach\footnote{We omit the
antisymmetric color wave function and the plane wave for the center of
mass motion which are common to all cases.}
\begin{itemize}
\item {\it $\Xi_{h_1h_2},\,\Omega_{h_1h_2}-$type baryons: }
\begin{eqnarray}
|\Xi_{h_1h_2},\,\Omega_{h_1h_2}; J=\frac12, M_J  \rangle  &=& 
\sum_{M_{S_h}M_{S_q}} ( 1\frac12\frac12|M_{S_h}M_{S_q}M_J)\  |h_1h_2;1M_{S_h}
 \rangle \otimes |q;\frac12 M_{S_q}\rangle \nonumber\\
&&\hspace{3cm}\times\Psi_{h_1h_2}^{\Xi,\,\Omega}(r_1,r_2,r_{12})
\end{eqnarray}
where $M_J$ is the third component of the baryon total angular
momentum while $|h_1h_2;S_h,M_{S_h}\rangle$ and $|q;\frac12
M_{S_q}\rangle $ represent spin states of the $h_1h_2$ subsystem and
the light quark respectively. $(j_1j_2j|m_1m_2m)$ is a Clebsch-Gordan
coefficient.  For $h_1=h_2$ we need
$\Psi_{h_1h_1}^{\Xi,\,\Omega}(r_1,r_2,r_{12})
=\Psi_{h_1h_1}^{\Xi,\,\Omega}(r_2,r_1,r_{12})$ to guarantee a complete
symmetry of the wave function under the exchange of the two heavy
quarks.
\item {\it $\Xi^*_{h_1h_2},\,\Omega^*_{h_1h_2}-$type baryons: }
\begin{eqnarray}
|\Xi^*_{h_1h_2},\,\Omega^*_{h_1h_2}; J=\frac32, M_J  \rangle  &=& 
\sum_{M_{S_h}M_{S_q}} ( 1\frac12\frac32|M_{S_h}M_{S_q}M_J)\  |h_1h_2;1M_{S_h}
 \rangle \otimes |q;\frac12 M_{S_q}\rangle \nonumber\\
&&\hspace{3cm}\times\Psi_{h_1h_2}^{\Xi^*,\,\Omega^*}(r_1,r_2,r_{12})
\end{eqnarray}
Similarly to the case before for $h_1=h_2$  we need $\Psi_{h_1h_1}
^{\Xi^*,\,
\Omega^*}(r_1,r_2,r_{12})
=\Psi_{h_1h_1}^{\Xi^*,\,\Omega^*}(r_2,r_1,r_{12})$. 
\item {\it $\Xi'_{h_1h_2},\,\Omega'_{h_1h_2}-$type baryons: }
\begin{eqnarray}
|\Xi'_{h_1h_2},\,\Omega'_{h_1h_2}; J=\frac12, M_J  \rangle  &=& 
  |h_1h_2;00 \rangle \otimes |q;\frac12 M_{J}\rangle \nonumber\\
&&\hspace{3cm}\times\Psi_{h_1h_2}^{\Xi',\,\Omega'}(r_1,r_2,r_{12})
\end{eqnarray}
In this case $h_1\ne h_2$ and we do not need the orbital part to have a 
definite symmetry under the exchange of the two quarks. 
\end{itemize}

The spatial wave functions $\Psi(r_1,r_2,r_{12})$ in the above
expressions will be determined by the variational principle: $\delta
\langle B | H^{\rm int}| B \rangle = 0$. For simplicity, we shall
assume a Jastrow--type functional form\footnote{ A similar form lead
to very good results in the case of baryons with a single heavy
quark~\cite{conrado04}.}:

\begin{eqnarray}
\Psi_{h_1h_2}^{B} (r_1,r_2,r_{12}) &=& N\, F^{B}(r_{12})\,
\phi_{h_1q}(r_1)\,\phi_{h_2q}(r_2)
\label{eq:wf}
\end{eqnarray}
where $N$  is a constant, which is determined from
normalization
\begin{eqnarray}
1=\int d^3r_1 \int d^3 r_2\ \left |\Psi_{h_1h_2}^{B}(r_1,r_2,r_{12})
\right |^2 = 8\pi^2 \int_0^{+\infty}dr_1~r_1^2~ \int_0^{+\infty}dr_2~r_2^2
\int_{-1}^{+1} d\mu~\left |\Psi_{h_1h_2}^{B}(r_1,r_2,r_{12})
\right |^2 
\end{eqnarray}
with $\mu$ being the cosine of the angle between the vectors $\vec
{r}_1$ and $\vec {r}_2$ ($r_{12}=(~r_1^2+r_2^2-2 r_1 r_2
\mu)^{1/2}$).
 
The functions $\phi_{h_1q}$ and $\phi_{h_2q}$ will be taken as the
$S-$wave ground states $\varphi_j(r_j)$ of the single particle
Hamiltonians $H^{sp}_{j}$ of Eq.~(\ref{eq:hint}) modified at large
distances.
\begin{eqnarray}
\phi_{h_jq} (r_j) &=& (1+\alpha_j\,r_j)\,\varphi_j(r_j),\ \ j=1,2
\label{eq:onebody}
\end{eqnarray}
The heavy--heavy correlation function $F^{B}$ will be given by a
linear combination of gaussians\footnote{Note that $F^{B}$ should
vanish at large distances because of the confinement potential. The
confinement potential is also responsible for the non--vanishing
values of the parameters $\alpha_j,\ j=1,2$ in
Eq.~(\protect\ref{eq:onebody}).}
\begin{eqnarray}
F^{B}(r_{12}) &=& 
\sum_{j=1}^4 a_j e^{-b_j^2(r_{12}+d_j)^2},\quad a_1=1 
\label{eq:f12}
\end{eqnarray}
The value of one of the $a_j$ parameters can be absorbed into the
normalization constant $N$, so that we fix $a_1=1$.  The rest of the
variational
parameters
are determined by the variational condition $\delta
\langle B | H^{\rm int}| B \rangle = 0$.

Although it is not self-evident from the functional form assumed, our
variational wave functions are consistent with the infinite heavy quark mass
limit, reducing in that limit to the product of the internal wave function for
the heavy diquark  times the wave function for the relative
motion of the light quark with respect to a pointlike heavy diquark. All this
is discussed in appendix~\ref{app:ihqml}.

\section{Static Properties}\label{sec:sp}

\subsection{Masses}
\label{subsec:masses}
The mass of the baryon is simply given by the expectation value of the
intrinsic Hamiltonian. Our results (VAR) appear in
Table~\ref{tab:xiomegamasses} where we compare them with the ones
obtained in Ref.~\cite{Si96} with the use of the same inter-quark
interactions but within a Faddeev approach (FAD). For that purpose we have 
eliminated from the latter
 a small three-body force contribution of the type 
$V_{123}= {\rm constant}/m_{h_1}m_{h_2}m_q$ that was also
included  in the evaluation of Ref.~\cite{Si96}. We will show their full 
results in the following tables.  Whenever comparison is
possible we find an excellent agreement between the two
calculations. In some cases the variational masses are even lower than
the Faddeev ones. Besides we give predictions for states that were not
considered in the study of Ref.~\cite{Si96}.
\begin{table}
\begin{tabular}{lcccccc}
& & AL1 & AL2 &AP1 &AP2 &BD  \\ \hline\tstrut

$\Xi_{cc}$\hspace{.5cm}      &VAR& 3612  & 3619  & 3629 &  3630 &  3639   \\
                &FAD~\cite{Si96}& 3609  & 3616  & 3625  &  3628 &  3633  \\ \\
		
$\Xi_{cc}^*$    &VAR& 3706  & 3715  & 3722  &  3729 &  3722  \\ \\
$ \Xi_{bb} $    &VAR& 10197 & 10180 & 10207 & 10179 & 10202  \\
                &FAD~\cite{Si96}& 10194 & 10175 & 10204 & 10176 & 10197  \\ \\
$\Xi_{bb}^*$    &VAR& 10236 & 10219 & 10245 & 10219 & 10235 \\ \\
$\Xi_{bc}$     &VAR& 6919  & 6912  & 6933  & 6917  & 6936   
\\ 
                &FAD~\cite{Si96}& 6916  & 6913  & 6928  & 6907  & 6934   \\ \\
$\Xi_{bc}'$      &VAR& 6948  & 6942  & 6957  & 6944  & 6965   \\\\
$\Xi_{bc}^*$   &VAR& 6986  & 6981  & 7000  & 6987  & 6993   \\ \hline

\end{tabular}\hspace{1cm}
\begin{tabular}{lc|ccccc}
 & & AL1 & AL2 &AP1 &AP2 &BD \\ \hline\tstrut

$\Omega_{cc}$ \hspace{.5cm}        &VAR&  3702 & 3718  & 3711 & 3710 & 3743  
\\
                     &FAD~\cite{Si96}&  3711 & 3718  & 3710 & 3709 & 3741\\ \\

$\Omega_{cc}^*$       &VAR&  3783 & 3802  & 3800 & 3802 & 3805 \\ \\
$\Omega_{bb}$         &VAR& 10260 & 10249 & 10259 &10226 &10274
\\
                      &FAD~\cite{Si96}& 10267 & 10246 & 10258 &10224 &10271\\
		      \\

$\Omega_{bb}^*$       &VAR& 10297 & 10287 & 10301 &10269 &10302\\ \\
$\Omega_{bc}$        &VAR& 6986  & 6986  & 6990 & 6969 & 7013 
\\
                      &FAD~\cite{Si96}&  7003 & 6996  & 6996 & 6971 & 7023 \\
		      \\

$\Omega_{bc}'$         &VAR& 7009  & 7010  & 7011 & 6994 & 7033 \\ \\
$\Omega_{bc}^*$      &VAR& 7046  & 7047  & 7055 & 7037 & 7057 \\ \hline %
\end{tabular}
\caption{Doubly heavy $\Xi$ and $\Omega$ baryons masses in MeV. VAR
stands for the results of our variational calculation. FAD stands for
the results obtained in Ref.~\cite{Si96} using the same interquark interactions
but within a Faddeev approach.}
\label{tab:xiomegamasses}
\end{table}

In Tables~\ref{tab:ximassesdiffmod} and \ref{tab:omegamassesdiffmod}
we compare our results with other theoretical calculations\footnote{Note in
Ref.~\cite{roncaglia95} the $\Xi_{bc},\,\Xi'_{bc}$ and 
$\Omega_{bc},\,\Omega'_{bc}$ baryons are defined such that the  total spin of the light 
$q$ quark and the heavy $c$ quark are well defined, being 0 for  
$\Xi_{bc},\,\Omega_{bc}$ and 1 for $\Xi'_{bc},\,\Omega'_{bc}$. They are thus  linear
combinations of our states. The different spin functions are related by
\begin{eqnarray}
\left(|qc;0\rangle\otimes|b;\frac12\rangle\right)^{J=1/2}=
\frac12\left(|bc;0\rangle\otimes|q;\frac12\rangle   \right)^{J=1/2}
-\frac{\sqrt3}{2}\left(|bc;1\rangle\otimes|q;\frac12\rangle  
\right)^{J=1/2}\nonumber\\
\left(|qc;1\rangle\otimes|b;\frac12\rangle\right)^{J=1/2}=
-\frac{\sqrt3}{2}\left(|bc;0\rangle\otimes|q;\frac12\rangle   \right)^{J=1/2}
-\frac{1}{2}\left(|bc;1\rangle\otimes|q;\frac12\rangle  
\right)^{J=1/2}
\end{eqnarray}
In order to extract their predictions  for the $\Xi_{bc},\,\Xi'_{bc}$ and 
$\Omega_{bc},\,\Omega'_{bc}$ baryons with total spin of the two heavy
quarks well defined, 
we have assumed that the above relations,  but with  coefficients square, are also valid for
the masses. Note this may be incorrect as we are neglecting a possible 
non negligible interference
contribution.}. Our
central values correspond to the results obtained with the AL1
potential, while the errors quoted take into account the variation
when using  different potentials. The same presentation is used for
the results obtained in Ref.~\cite{Si96} for which we now show their full
values including the contribution of the three-body force.  All calculations give
similar results that vary within a few per cent. From the experimental
point of view the SELEX Collaboration~\cite{mattson02} has recently
measured the value of $M_{\Xi_{cc}}$. This experimental value is 100
MeV smaller that our result. On account of what has been said in the
introduction, one should take this experimental value with due
caution. Note also that in
Ref.~\cite{mattson02}  the systematic
error is not given. There are also different lattice determinations for baryons with two 
equal heavy quarks~\cite{khan00,lewis01,flynn03}. Our results agree within errors with 
the lattice data for baryons with two $c$ quarks, while they are roughly 100\,MeV below lattice
results for doubly $b$-quark baryons. The best overall agreement with  lattice
data available so far is achieved in the calculation of Ref.~\cite{roncaglia95}
where they use the Feynman-Hellmann theorem and semiempirical mass formulas
in the framework of a nonrelativistic quark model but without the use of an
explicit Hamiltonian. Within full dynamical calculations, ours and the
relativistic calculations in Refs.~\cite{ebert02,tong00,ebert97} have the best
overall agreement with lattice data.
\begin{table}[h]
\begin{tabular}{lccccccccccccc}
& This work&\cite{Si96}&
\cite{ebert02}&\cite{kiselev02}&\cite{narodetskii02}&\cite{tong00}&\cite{itoh00}
&\cite{vijande04}&\cite{gershtein00}&\cite{ebert97}&\cite{roncaglia95}&\cite{korner94}&\cite{mathur02}\\\hline
$\Xi_{cc}$\hspace{.5cm}  &$3612^{+17}$
&$3607^{+24}$&3620&3480&3690&3740&3646&3524&3478&3660&$3660\pm70$&3610&$3588\pm72$\\   
$\Xi^*_{cc}$  &$3706^{+23}$ & &3727&3610&&3860&3733&3548&3610&3810&$3740\pm80$&3680&\\   
$\Xi_{bb}$  &$10197^{+10}_{-17}$&$10194^{+10}_{-19}$
&10202&10090&10160&10300&&&10093&10230&$10340\pm100$\\   
$\Xi^*_{bb}$  &$10236^{+9}_{-17}$&
&10237&10130&&10340&&&10133&10280&$10370\pm100$\\   
$\Xi_{bc}$  &$6919^{+17}_{-7}$ &$6915^{+17}_{-9}$ &6933&6820&6960&7010&&&6820
&6950&$6965\pm90^\dagger$
&&$6840\pm236$\\   
$\Xi'_{bc}$  &$6948^{+17}_{-6}$ &&6963&6850&&7070&&&6850&7000&$7065\pm90^\dagger$
\\   
$\Xi^*_{bc}$  &$6986^{+14}_{-5}$ &&6980&6900&&7100&&&6900&7020&$7060\pm90$ \\   
\end{tabular}\vspace{1cm}
\begin{tabular}{lcccccc}
& This work&Exp.~\cite{mattson02}
&Latt.~\cite{khan00}&Latt.~\cite{lewis01}&Latt.~\cite{flynn03}\\\hline
$\Xi_{cc}$\hspace{.5cm}  &$3612^{+17}$
&$3519\pm1$&&$3605\pm23$&$3549\pm 95$ \\   
$\Xi^*_{cc}$  &$3706^{+23}$ &  &&$3685\pm 23$&$3641\pm97$\\   
$\Xi_{bb}$  &$10197^{+10}_{-17}$&&$10314\pm47$\\   
$\Xi^*_{bb}$  &$10236^{+9}_{-17}$& &$10333\pm 55$\\   
\end{tabular}\caption{First panel: doubly heavy $\Xi$ masses in MeV as obtained in different
models. Our central values, and the ones of Ref.~\cite{Si96}, have
been evaluated with the AL1 potential. Second panel: we compare our results with 
the experimental
value for $M_{\Xi_{cc}}$ measured by the SELEX
Collaboration~\cite{mattson02} (Note the cautions that appear on this 
experimental mass in the introduction), and  lattice results from 
Refs.~\cite{khan00,lewis01,flynn03}. Entries with $^\dagger$ should be taken
with due caution (see footnote 6). }

\label{tab:ximassesdiffmod}
\end{table}

\begin{table}[h!!!]
\begin{tabular}{lcccccccccccc}
& This work&\cite{Si96}&
\cite{ebert02}&\cite{kiselev02}&\cite{narodetskii02}&\cite{tong00}&\cite{itoh00}
&\cite{gershtein00}&\cite{ebert97}&\cite{roncaglia95}&\cite{korner94}&\cite{mathur02}
\\\hline
$\Omega_{cc}$\hspace{.5cm}  &$3702^{+41}$  &$3710^{+29}_{-2}$& 3778&3590&3860&3760&3749&
3590&3760&$3740\pm70$&3710&$3698\pm65$\\   
$\Omega^*_{cc}$ &$3783^{+22}$ &&  3872&3690&&3900&3826&3690&3890&$3820\pm80$&3760\\   
$\Omega_{bb}$  &$10260^{+14}_{-34}$ &$10267^{+4}_{-43}$&
10359&10180&10340&10340&&10180&10320&$10370\pm100$&\\   
$\Omega^*_{bb}$ &$10297^{+5}_{-28}$ &&
10389&10200&&10380&&10200&10360&$10400\pm100$&\\   
$\Omega_{bc}$   &$6986^{+27}_{-17}$
&$7003^{+20}_{-32}$&7088&6910&7130&7050&&6910&7050&$7045\pm90^\dagger$
&&$6954\pm225$\\   
$\Omega'_{bc}$  &$7009^{+24}_{-15}$& &7116&6930&&7110&&6930&7090&$7105\pm90^\dagger$
&\\   
$\Omega^*_{bc}$ &$7046^{+11}_{-9}$ &&7130&6990&&7130&&6990&7110&$7120\pm90$& \\   
\end{tabular}\vspace{1cm}
\begin{tabular}{lcccc}
& This work&Latt.~\cite{khan00}&Latt.~\cite{lewis01}&Latt.~\cite{flynn03}\\\hline
$\Omega_{cc}$\hspace{.5cm}  &$3702^{+41}$  &&$3733\pm9^{+7}_{-38}$&$3663\pm 97$\\   
$\Omega^*_{cc}$ &$3783^{+22}$ &&$3801\pm9^{+3}_{-34}$&$3734\pm 98$ \\   
$\Omega_{bb}$  &$10260^{+14}_{-34}$ &$10365\pm40^{-11}_{+12}$$^{+16}_{-0}$\\   
$\Omega^*_{bb}$ &$10297^{+5}_{-28}$ &$10383\pm39^{-8}_{+8}$$^{+12}_{-0}$\\   
\end{tabular}\caption{Same as Table~\ref{tab:ximassesdiffmod} for doubly heavy $\Omega$
baryons.}

\label{tab:omegamassesdiffmod}
\end{table}

There are also independent determinations of mass splittings in lattice
QCD~\cite{khan00,lewis01,flynn03}, nonrelativistic lattice QCD~\cite{mathur02}
and pNRQCD~\cite{brambilla05}. In Table~\ref{tab:ms} we compare those results to
the ones obtained in the present calculation and in other models.
\begin{table}
\begin{tabular}{lccccccccccc}
& This work&\cite{ebert02}&\cite{kiselev02}&\cite{tong00}&\cite{gershtein00}&\cite{ebert97}&\cite{brambilla05}&\cite{mathur02}&Latt.~\cite{khan00}&Latt.~\cite{lewis01}&Latt.~\cite{flynn03}\\\hline
$M_{\Xi^*_{cc}}-M_{\Xi_{cc}}$\hspace{.5cm}&$94^{+5}_{-11}$&107&130&120&132&150&$120\pm40$&$70\pm13$&&$80\pm11$&$87\pm 19$  \\ 
$M_{\Xi^*_{bb}}-M_{\Xi_{bb}}$\hspace{.5cm}&$39^{+1}_{-6}$&35&40&40&40&50&$34\pm4$&$20\pm7$&$20\pm6$\\ 
$M_{\Xi^*_{bc}}-M_{\Xi_{bc}}$\hspace{.5cm}&$67^{+3}_{-10}$&47&80&90&80&70&&$43\pm11$\\ 
$M_{\Xi'_{bc}}-M_{\Xi_{bc}}$\hspace{.5cm}&$29^{+1}_{-5}$&30&30&60&30&50&&$9\pm7$\\ 
$M_{\Omega^*_{cc}}-M_{\Omega_{cc}}$\hspace{.5cm}&$81^{+11}_{-19}$&94&100&140&100&130&&$63\pm9$&&$68\pm7$&$67\pm 16$  \\   
$M_{\Omega^*_{bb}}-M_{\Omega_{bb}}$\hspace{.5cm}&$37^{+6}_{-9}$&30&20&40&20&40&&$19\pm5$&$20\pm5$  \\   
$M_{\Omega^*_{bc}}-M_{\Omega_{bc}}$\hspace{.5cm}&$60^{+8}_{-16}$&42&80&80&80&60&&$39\pm8$  \\   
$M_{\Omega'_{bc}}-M_{\Omega_{bc}}$\hspace{.5cm}&$23^{+2}_{-3}$&28&20&60&20&40&&$9\pm6$  \\   
\end{tabular}
\caption{Mass splittings in MeV for doubly heavy $\Xi$ and $\Omega$
baryons. Our central values have been obtained with the AL1 potential. Entries
with a $^\dagger$ should be taken
with due caution (see footnote 6).}
\label{tab:ms}
\end{table}
Our central results evaluated with the AL1 potential are larger than the ones obtained in lattice QCD~\cite{khan00,lewis01,flynn03} 
and lattice nonrelativistic QCD~\cite{mathur02}. The agreement is better when
we use  the BD potential of
Ref.~\cite{BD81} for
which we  always get the lowest results. Similar results are obtained by 
the relativistic calculation of
Ref.~\cite{ebert02}, whereas for the  relativistic
calculations in Refs.~\cite{kiselev02,tong00,ebert97} and the nonrelativistic
one in Ref.~\cite{gershtein00} the
agreement  with lattice QCD and  nonrelativistic lattice QCD data   worsens. As
for the calculation in Ref.~\cite{roncaglia95} we do not quote their results due
to the large theoretical errors involved.
\subsection{Charge  densities and radii}
\label{subsec:chmdr}
The baryon charge density at the point $P$ (coordinate vector $\vec{r}$ in 
the CM frame, see Fig.~\ref{fig:coor}) is given by: 
\begin{eqnarray}
\rho_e^{B} (\vec{r}\,) 
& = & \int d^3 r_1 d^3 r_2\ \Big |
\Psi_{h_1h_2}^{B}(r_1,r_2,r_{12}) \Big |^2 \left \{ e_{h_1} \delta^3
(\vec{r}-\vec{y}_{h_1}) + e_{h_2} \delta^3 (\vec{r}-\vec{y}_{h_2}) + e_{q}
\delta^3 (\vec{r}-\vec{y}_q) \right \} \nonumber \\ 
& \equiv & \rho_e^{B} (\vec{r}\,)\big|_{h_1} 
+ \rho_e^{B}(\vec{r}\,)\big|_{h_2}  +
\rho_e^{B}(\vec{r}\,)\big|_{q} \label{eq:dens}
\end{eqnarray} 
where $e_{h_1,\,h_2,\,q}$ are the quark charges in proton charge units
$e$, and from Fig.~\ref{fig:coor} we have\footnote{There exists the
obvious restriction
$m_{h_1}\vec{y}_{h_1}+m_{h_2}\vec{y}_{h_2}+m_{q}\vec{y}_q=\vec 0$.}
$\vec{y}_{h_1}=\vec{y}_q+\vec{r}_1$,
$\vec{y}_{h_2}=\vec{y}_q+\vec{r}_2$ and $\vec{y}_q= - \left
(m_{h_1}\vec{r}_1+m_{h_2}\vec{r}_2 \right)/\overline M$.  Since our
$L=0$ wave functions only depend on scalars ($r_1,r_2$ and $r_{12}$)
the charge density is spherically symmetric ($\rho_e^{B} (\vec{r}\,) =
\rho_e^{B} (|\vec{r}~|)$).

The charge form factor is defined as usual
\begin{eqnarray}
{\cal F}_e^{B} (\vec{q}~) = \int d^3 r\ e^{{\rm
i}\vec{q}\cdot\vec{r}}\rho_e^{B} (r)\label{eq:fe}
\end{eqnarray}
and it only depends on $|\vec{q}~|$. Its value at $\vec{q}=\vec 0$
 gives the baryon charge in units of the proton charge.

\begin{figure}
\resizebox{13.cm}{!}{\includegraphics{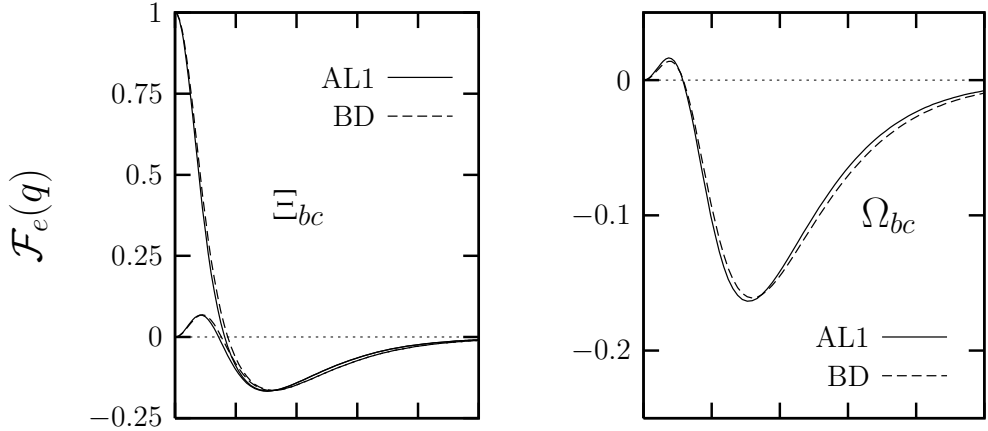}}
\caption{Charge form factor of the $\Xi_{bc}$
and $\Omega_{bc}$ baryons evaluated with the AL1
(solid line) and BD (dashed line) potentials. For  $\Xi_{bc}$ baryons we show the
two possible charge states.
Similar results are obtained for $\Xi^*_{bc},\,\Xi'_{bc}$ and 
$\Omega^*_{bc},\,\Omega'_{bc}$
 }\label{fig:cffbc}
\end{figure}

The charge mean square radii are defined
\begin{equation}
\langle r^2 \rangle_e^{B} = 
\int d^3 r\ r^2 \rho_e^{B} (r) = 
4\pi \int_0^{+\infty}dr\ r^4 \rho_e^{B} (r) 
\label{eq:r2q}
\end{equation}
\begin{figure}
\resizebox{13.cm}{!}{\includegraphics{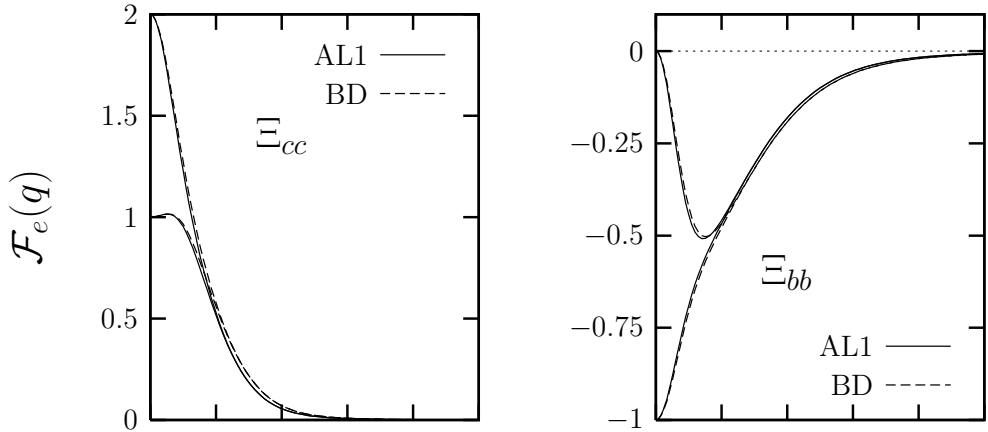}}\vspace{.5cm}\\
\caption{ Charge form factor of the $\Xi_{cc}$ and $\Xi_{bb}$
 baryons evaluated with the AL1~\cite{Si96,SS94}
(solid line) and BD~\cite{BD81} (dashed line) potentials. We show the two
possible charge states. Similar results are obtained for $\Xi^*_{cc}$ and
$\Xi^*_{bb}$.}
\label{fig:cffxi}
\end{figure}
\begin{figure}
\vspace{-1cm}
\resizebox{13.cm}{!}{\includegraphics{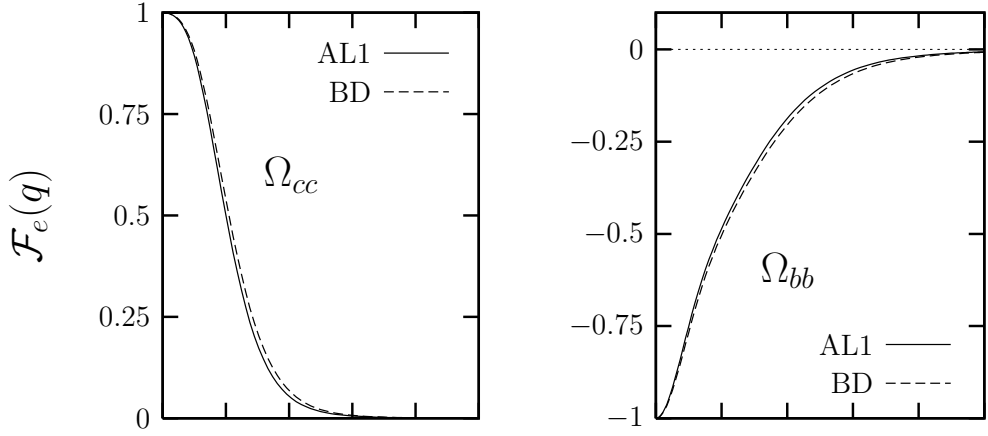}}
\caption{ Charge form factor of the 
 $\Omega_{cc}$ and $\Omega_{bb}$  baryons 
 evaluated with the AL1 (solid line)
and BD (dashed line) potentials. Similar results are obtained for 
$\Omega^*_{cc}$ and $\Omega^*_{bb}$.
}\label{fig:cffomega}
\end{figure}
In Figs.~\ref{fig:cffxi}, \ref{fig:cffomega} and \ref{fig:cffbc} we
show the charge form factors for $\Xi_{cc},\, \Xi_{bb}$, $\Omega_{cc},\,\Omega_{bb}$
and $\Xi_{bc},\,\Omega_{bc}$ baryons. In each case similar results are 
obtained for the star and prime excitations. We show the calculations with both the AL1 potential
of Refs.~\cite{Si96,SS94} and the BD potential of
Ref.~\cite{BD81}. The differences between the two calculations are
minor in most cases.

In Table~\ref{tab:xiomegacr} we show the charge mean square
radii. With the exceptions of the $\Xi_{bc}^0$ and $\Omega_{bc}^0$, we
find good agreement with the results obtained in Ref.~\cite{Si96}
within a Faddeev calculation. The possible presence of a $S_h=0$
contribution in the wave functions of Ref.~\cite{Si96} could be the
possible explanation for this discrepancy. We also compare with the
results obtained, for a few states, in Ref.~\cite{bruno04} with the
use a relativistic quark model in the instant form. The agreement is
bad in this case.
\begin{table}[h!!!]
\begin{tabular}{lccc}
&This work & \cite{Si96}&\cite{bruno04}\\ \hline

$\Xi_{cc}^+$\hspace{.5cm}   &$-0.030^{+0.003}_{-0.016}$ &$-0.038^{+0.004}_{-0.016}$&$0.1$ \\
$\Xi_{cc}^{++}$ & $0.298^{+0.034}_{-0.028}$  &$0.315^{+0.035}_{-0.030}$&$1.3$ \\
		 \\
$\Xi_{cc}^{*+}$    &$-0.042^{+0.007}_{-0.019}$& & \\
$\Xi_{cc}^{*++}$    &$0.341^{+0.041}_{-0.042}$&&\\ \\
$ \Xi_{bb}^0 $    &$0.221^{+0.033}_{-0.025}$& $0.242^{+0.035}_{-0.027}$&\\
$ \Xi_{bb}^- $    &$-0.133^{+0.014}_{-0.016}$& $-0.143^{+0.006}_{-0.018}$&\\ \\
$\Xi_{bb}^{*-}$    &$-0.142^{+0.018}_{-0.018}$& &\\ \\
$\Xi_{bb}^{*0}$    &$0.238^{+0.035}_{-0.031}$& &\\ \\
$\Xi_{bc}^0$     &$-0.057^{+0.006}_{-0.013}$& $-0.072^{+0.008}_{-0.017}$&\\  
$\Xi_{bc}^+$     &$0.279^{+0.026}_{-0.031}$& $0.306^{+0.035}_{-0.011}$&\\ \\
$\Xi_{bc}'^0$      &$-0.065^{+0.010}_{-0.015}$& & \\
$\Xi_{bc}'^+$      &$0.283^{+0.036}_{-0.025}$& & \\
\\
$\Xi_{bc}^{*0}$   &$-0.065^{+0.010}_{-0.018}$& &\\ 
$\Xi_{bc}^{*+}$   &$0.305^{+0.031}_{-0.039}$&& \\ 
\hline
\end{tabular}\hspace{0.9cm}
\begin{tabular}{lccc}
 &This work &\cite{Si96}&\cite{bruno04} \\ \hline 

$\Omega_{cc}^+$\hspace{0.5cm} &$0.013^{+0.001}_{-0.002}$&$0.009_{-0.003}$&$0.2$\\ \\

$\Omega_{cc}^{*+}$       &$0.009^{+0.001}_{-0.002}$& & \\ \\
$\Omega_{bb}^-$         &$-0.086^{+0.008}_{-0.001}$&   $-0.090^{+0.007}_{-0.002}$&\\ \\

$\Omega_{bb}^{*-}$       &$-0.092^{+0.011}_{-0.001}$& &\\ \\
$\Omega_{bc}^0$        &$-0.016_{+0.003}$& $-0.025^{+0.002}_{-0.003}$& \\\\

$\Omega_{bc}'^0$         &$-0.019^{+0.003}_{-0.003}$&&\\ \\
$\Omega_{bc}^{*0}$      &$-0.021^{+0.004}_{-0.002}$& &\\ \hline %
\vspace{2.7cm}

\end{tabular}
\caption{Charge mean square radii in fm$^2$ for doubly heavy $\Xi$ and
 $\Omega$ baryons.  Our central values, and the ones of
 Ref.~\cite{Si96}, have been evaluated with the AL1 potential.}
\label{tab:xiomegacr}
\end{table}

\subsection{Magnetic moments}
\label{subsec:mm}
The orbital part of the magnetic moment is defined in terms of the
velocities $\vec{v}$ of the quarks, with respect to the position of
the CM, and it reads
\begin{eqnarray}
{\mu}^{B} &=& \int  d^3 r_1 d^3 r_2 \left(\Psi_{h_1h_2}^{B}(r_1,r_2,r_{12})\right)^* \left \{
\frac{e_{h_1}}{2m_{h_1}} (\vec{y}_{h_1} \times m_{h_1}\vec{v}_{{h_1}})_z \right.\nonumber \\
&+&\left. \frac{e_{h_2}}{2m_{h_2}} (\vec{y}_{h_2} \times
m_{h_2}\vec{v}_{h_2})_z + 
\frac{e_{q}}{2m_{q}} (\vec{y}_q \times m_{q}
\vec{v}_{y_q})_z \right \}\Psi_{h_1h_2}^{B}(r_1,r_2,r_{12})
\end{eqnarray}
with\footnote{Note that the classical kinetic energy  has a term on 
$\vec{v}_{h_1}\cdot\vec{v}_{h_2}$ and then the operator
$m_{h_1} \vec{v}_{h_1}$ is not proportional to 
$-i\stackrel{\to}{\nabla}_{y_{h_1}}$, but it is rather given by
$m_{h_1} \vec{v}_{h_1}={(\overline M-m_{h_1})}/{\overline M}\cdot
(-i\stackrel{\to}{\nabla}_{y_{h_1}}) 
-{m_{h_2}}/{\overline M}\cdot(-i\stackrel{\to}{\nabla}_{y_{h_2}})
=(-i\stackrel{\to}{\nabla}_1)$. Similarly
 $m_{h_2} \vec{v}_{h_2}=(-i\stackrel{\to}{\nabla}_2)$.}
 $m_{h_1}\vec{v}_{h_1} = - i \vec{\nabla}_1$,
$m_{h_2}\vec{v}_{h_2} = - i \vec{\nabla}_2$ and $m_q\vec{v}_{q}
= i \left ( \vec{\nabla}_1 + \vec{\nabla}_2 \right)$. Since our orbital wave
function has $L=0$, the orbital
 magnetic moment vanishes. The magnetic moment of the baryon is then
 entirely given by the spin contribution.
\begin{eqnarray}
\langle B;\, J,\,M_J=J|\ \frac{e_{h_1}}{2m_{h_1}} (\vec{\sigma}_{h_1})_z
+ \frac{e_{h_2}}{2m_{h_2}} (\vec{\sigma}_{h_2})_z  + 
\frac{e_{q}}{2m_{q}} (\vec{\sigma}_q)_z\ | B;\,J,\, M_J=J\rangle
\end{eqnarray}
 Those matrix elements are trivially evaluated with the
results
\begin{eqnarray}
\Xi_{h1h_2},\ \Omega_{h1h_2}&\longrightarrow&
\frac{2}{3}\left( \frac{e_{h_1}}{2m_{h_1}}+\frac{e_{h_2}}{2m_{h_2}}-\frac12
\ \frac{e_{q}}{2m_{q}} \right)\nonumber\\
\Xi^*_{h1h_2},\ \Omega^*_{h1h_2}&\longrightarrow&
 \frac{e_{h_1}}{2m_{h_1}}+\frac{e_{h_2}}{2m_{h_2}}+
\frac{e_{q}}{2m_{q}} \nonumber\\
\Xi'_{h1h_2},\ \Omega'_{h1h_2}&\longrightarrow&
 \frac{e_{q}}{2m_{q}}
\end{eqnarray}

In Table~\ref{tab:xiomegamm} we give our numerical results. Our
central values, as the ones obtained in Ref.~\cite{Si96} within a
Faddeev approach, have been evaluated with the use of the AL1
potential. When compared to the values obtained in Ref.~\cite{Si96} we
find very good agreement with just a few exceptions
($\Xi_{bc}^0,\,\Xi_{bc}^+,\,\Omega_{bc}^0$). The discrepancy for the
latter baryons may come from a possible non negligible $S_h=0$
contribution to their wave functions in the calculation of
Ref.~\cite{Si96}. In our case we have fixed $S_h=1$ which we think is
a very good approximation since in the limit of infinite heavy quark
masses the spin of the heavy quark degrees of freedom is well defined.

We also compare our results to the ones obtained in
Refs.\cite{lichtenberg77,faessler06,bruno04,oh91,jena86,bose80} using
different approaches. The differences between different calculations are
 in some cases large. Being
$L=0$ a good approximation the magnetic
moments are essentially determined by the spin contribution of the
quarks. With $m_b\gg m_u,\,m_d,\,m_s$, the contribution from the $b$ quarks is
negligible compared to the one of the light quark. This is also true to a lesser
extent for the $c$ quark.
\begin{table}[h!!!]
\begin{tabular}{lcccccccc}
& This work& \cite{Si96}&\cite{lichtenberg77}&\cite{faessler06}&\cite{bruno04}&\cite{oh91}
&\cite{jena86}&\cite{bose80} \\ \hline

$\Xi_{cc}^+$\hspace{.5cm} &$0.785^{+0.050}_{-0.030}$ &$0.784^{+0.050}_{-0.029}$&0.806
&0.72&$0.72$&$0.89\sim0.98$&$0.778\sim0.790$&0.86\\
$\Xi_{cc}^{++}$&$-0.208^{+0.035}_{-0.086}$
&$-0.206^{+0.034}_{-0.086}$&$-0.124$&0.13&$-0.10$&$-0.47$&$-0.172\sim-0.154$&0.17\\
		 \\
$\Xi_{cc}^{*+}$ &$-0.311^{+0.052}_{-0.130}$ &$$ &$-0.186$&
&&$-1.17\sim-0.98$&&0.20\\
$\Xi_{cc}^{*++}$&$2.67^{+0.27}_{-0.15}$ &$$ &2.60 &&&$3.16\sim3.18$&&2.54\\ \\
$ \Xi_{bb}^0 $
&$-0.742^{+0.044}_{-0.091}$&$-0.742^{+0.044}_{-0.092}$&&$-0.53$&&&$-0.726\sim-0.705$&0.61\\
$ \Xi_{bb}^- $  &$0.251^{+0.045}_{-0.021}$ &$0.251^{+0.046}_{-0.021}$ &&0.18
&&&$0.226\sim0.236$&0.14\\ \\
$\Xi_{bb}^{*0}$ &$1.87^{+0.27}_{-0.13}$  &$$ && &&&&1.37\\ 
$\Xi_{bb}^{*-}$ &$-1.11^{+0.06}_{-0.14}$ &$$ &&& &&&$-0.95$\\  \\
$\Xi_{bc}^0$
&$0.518^{+0.048}_{-0.020}$&$0.058^{+0.059}_{-0.054}$&& 0.42   &&&&\\
$\Xi_{bc}^+$  &$-0.475^{+0.040}_{-0.088}$  &$-0.198^{+0.057}_{-0.056}$ &&$-0.12$
&&&& \\  \\

$\Xi_{bc}'^0$ &$-0.993^{+0.065}_{-0.137}$  &$$ &&$-0.76$&&& $-0.385\sim-0.366$ & \\
$\Xi_{bc}'^+$ &$1.99^{+0.27}_{-0.13}$  &$$ && 1.52 & &&  $1.50\sim1.54$  &\\
\\
$\Xi_{bc}^{*0}$ &$-0.712^{+0.059}_{-0.133}$  &$$ &&&&&   &$-0.39$\\ 
$\Xi_{bc}^{*+}$ &$2.27^{+0.27}_{-0.14}$ &$$ &&&&&  &2.04\\ 
\hline
\end{tabular}\vspace{1cm}\\
\begin{tabular}{lcccccccc}
 & This work&\cite{Si96}&\cite{lichtenberg77}&\cite{faessler06}&\cite{bruno04}&\cite{oh91} 
& \cite{jena86}&\cite{bose80}\\ \hline 

$\Omega_{cc}^+$\hspace{.5cm}  &$0.635^{+0.012}_{-0.015}$ &$0.635^{+0.011}_{-0.015}$ &0.688&0.67
&$0.72$&$0.59\sim0.64$&$0.657\sim0.663$&0.84\\ \\

$\Omega_{cc}^{*+}$  &$0.139^{+0.009}_{-0.017}$   &$$ &0.167&
&&$-0.20\sim0.03$&&0.39\\ \\
$\Omega_{bb}^-$ &$0.101^{+0.007}_{-0.007}$ &$0.101^{+0.007}_{-0.006}$ &&0.04& &&
$0.105\sim0.108$&0.084\\ \\

$\Omega_{bb}^{*-}$  &$-0.662^{+0.022}_{-0.024}$  &$$&  & &&&&$-1.28$\\ \\
$\Omega_{bc}^0$ &$0.368^{+0.010}_{-0.011}$&$0.009^{+0.038}_{-0.029}$ &&0.45&&&&\\
		      \\

$\Omega_{bc}'^0$  &$-0.542^{+0.021}_{-0.024}$   &$$&
&$-0.61$&&& $-0.130\sim-0.125$    &\\ \\
$\Omega_{bc}^{*0}$ &$-0.261^{+0.015}_{-0.021}$  &$$& &&&&&$-0.22$\\ \hline %

\end{tabular}
\caption{Magnetic moments, in nuclear magnetons ($|e|/2m_p$, with
  $m_p$ the proton mass), of doubly heavy $\Xi$ and $\Omega$
  baryons. Our central values, and the ones of Ref.~\cite{Si96}, have
  been evaluated with the AL1 potential.}
\label{tab:xiomegamm}
\end{table}

\section{Semileptonic decay}
\label{sec:sd}
In this section we shall use the wave functions obtained with the
variational method to study different doubly $B(1/2^+)\to B'(1/2^+)$
baryon semileptonic decays involving a $b\to c $ transition at the
quark level.

The differential decay width reads
\begin{equation}
{\rm d}\Gamma= 
8 |V_{cb}|^2 m_{B'} G_F^{\,2}  
 \frac{d^3p^\prime}{(2\pi)^32E^\prime_{B'} }
\frac{d^3k}{(2\pi)^32E_{\bar \nu_l} } \frac{d^3k^\prime}{(2\pi)^32E^\prime_{l}
}  (2\pi)^4 \delta^4(p-p^\prime-k-k^\prime)\ {\cal
L}^{\alpha\beta}(k,k')
{\cal H}_{\alpha\beta}(p,p')    
\end{equation}
where $|V_{cb}|$ is the modulus of the corresponding
Cabibbo--Kobayashi--Maskawa matrix element, $m_{B'}$ is the mass of
the final baryon, $G_F= 1.16637(1)\times
10^{-11}$\,MeV$^{-2}$\cite{pdg06} is the Fermi decay constant,
$p$, $p'$, $k$ and $k'$ are the four-momenta of the initial baryon,
final baryon, final anti-neutrino and final lepton respectively, and
${\cal L}$ and ${\cal H}$ are the lepton and hadron tensors.

The lepton tensor
is given as
\begin{eqnarray}
{\cal L}^{\mu\sigma}(k,k')&=& k'^\mu k^\sigma +k'^\sigma k^\mu
- g^{\mu\sigma} k\cdot k^\prime + {\rm i}
\epsilon^{\mu\sigma\alpha\beta}k'_{\alpha}k_\beta \label{eq:lep}
\end{eqnarray}
where we use the convention $\epsilon^{0123}=-1$, $g^{\mu\mu}=(+,-,-,-)$.

The hadron tensor is given as
\begin{eqnarray}
{\cal H}_{\mu\sigma}(p,p') &=& \frac12 \sum_{r,r'}  
 \left\langle B', r'\
\vec{p}^{\,\prime}\left|\,
\overline \Psi^c(0)\gamma_\mu(I-\gamma_5)\Psi^b(0)\right| B, r\ \vec{p}   \right\rangle 
\ \left\langle B', r'\ 
\vec{p}^{\,\prime}\left|\,\overline \Psi^c(0)\gamma_\sigma(I-\gamma_5)
\Psi^b(0) \right|  B, r\ \vec{p} \right\rangle^*
\label{eq:wmunu}
\end{eqnarray}
with $\left|B, r\ \vec p\right\rangle\, (\left|B', r'\
\vec{p}\,'\right\rangle)$ representing the initial (final) baryon with
three--momentum $\vec p$ ($\vec{p}\,'$) and spin third component $r$
($r'$). The baryon states are normalized such that $\langle r\
\vec{p}\, |\, r' \ \vec{p}^{\,\prime} \rangle = (2\pi)^3 (E(\vec
p\,)/m)\,\delta_{rr'}\, \delta^3(\vec{p}-\vec{p}^{\,\prime})$.  The
hadron matrix elements can be parametrized in terms of six form
factors as
\begin{eqnarray}
\left\langle B', r'\ \vec{p}^{\,\prime}\left|\,\overline \Psi^c(0)\gamma_\mu(I-\gamma_5)\Psi^b(0)
 \right| B, r\ \vec{p}
\right\rangle& =& {\bar u}^{B'}_{r'}(\vec{p}^{\,\prime})\Big\{
\gamma_\mu\left(F_1(w)-\gamma_5 G_1(w)\right)+ v_\mu\left(F_2(w)-\gamma_5
G_2(w)\right)\nonumber\\
&&\hspace{1.5cm}+v^\prime_\mu\left(F_3(w)-\gamma_5 G_3(w)
\right)\Big\}u^{B}_r(\vec{p}\,) \label{eq:def_ff}
\end{eqnarray}
where $u^{B,B'}$ are dimensionless Dirac spinors, normalized as ${\bar
u} u = 1$, and $v_\mu = p_\mu/m_{B}$ ($v^\prime_\mu =
p^\prime_\mu/m_{B'}$) is the four velocity of the initial $B$ (final
$B'$) baryon. The form factors are functions of the velocity transfer
$w=v\cdot v^\prime$ or equivalently of the four momentum transfer
($q=p-p'$) square $q^2= m_{B}^2 + m_{B'}^2 - 2m_{B}m_{B'}w$. In the
decay $w$ ranges from $w=1$, corresponding to zero recoil of the final
baryon, to a maximum value given by $w=w_{\rm max}= (m_{B}^2 +
m_{B'}^2)/(2m_{B}m_{B'})$ which depends on the transition.

Neglecting lepton masses, we have for the differential decay rates
from transversely $(\Gamma_T)$ and longitudinally $(\Gamma_L)$
polarized $W$'s (the total width is
$\Gamma=\Gamma_L+\Gamma_T$)~\cite{korner92}
\begin{eqnarray}
\frac{{\rm d}\Gamma_T}{{\rm d}w}&=&
\frac{G^2_F |V_{cb}|^2}{12\pi^3}m_{B'}^3\sqrt{w^2-1}\,q^2 
\Big\{ (w-1)|F_1(w)
|^2+(w+1)|G_1(w)|^2 \Big\} \nonumber\\
&&\nonumber\\
\frac{{\rm d}\Gamma_L}{{\rm d}w}&=&
\frac{G^2_F |V_{cb}|^2}{24\pi^3}m_{B'}^3\sqrt{w^2-1}
\Big\{(w-1)|{\cal F}^V(w)|^2 + (w+1)|{\cal
F}^A(w)|^2  \Big\} \nonumber\\
&&\nonumber\\
 {\cal F}^{V,A}(w) &=& \Big[ (m_{B}\pm m_{B'}) F_1^{V,A}(w) +
(1\pm w)\left(m_{B'} F_2^{V,A}(w)+m_{B}
F_3^{V,A}(w)\right)\Big],\nonumber\\&& \quad F_j^V \equiv F_j(w) , ~ F_j^A \equiv
G_j(w),~ j=1,2,3\nonumber\\
 \label{eq:dg}
\end{eqnarray}

One can also evaluate the polar angle distribution~\cite{korner92}:
\begin{equation}
\frac{{\rm d}^2\Gamma}{{\rm d}w\,{\rm d}\cos\theta} = \frac38
\left(\frac{{\rm d}\Gamma_T}{{\rm d}w} +
 2\frac{{\rm d}\Gamma_L}{{\rm d}w} \right)\Big\{1+2\alpha^\prime\cos\theta +
 \alpha^{\prime\prime} \cos^2\theta \Big\}
\label{eq:asymmetry1}
\end{equation}
where $\theta$ is the angle between $\vec{k}^\prime$ and
$\vec{p}^{\,\prime}$ measured in the off--shell $W$ rest frame, and
$\alpha^\prime$ and $\alpha^{\prime\prime}$ are asymmetry parameters
given by
\begin{eqnarray}
\alpha^\prime &=&  - \frac{G^2_F
    |V_{cb}|^2}{6\pi^3} 
{m_{B'}^3}\frac{
    q^2\,(w^2-1)\,F_1(w)G_1(w)}{{\rm d}\Gamma_T/{{\rm d}w}+
    2\,{\rm d}\Gamma_L/{{\rm d}w}}  \nonumber\\
&&\nonumber\\
\alpha^{\prime\prime} &=& \frac{{\rm d}\Gamma_T/{{\rm d}w}  
   - 2\,{\rm d}\Gamma_L/{{\rm d}w}}
   {{\rm d}\Gamma_T/{{\rm d}w}+
    2\,{\rm d}\Gamma_L/{{\rm d}w}} 
\label{eq:asymmetry2}
\end{eqnarray} 
These asymmetry parameters  are functions of the velocity
transfer $w$ and on averaging over $w$ the numerators and denominators
are integrated separately and  thus we have
\begin{eqnarray}
\langle \alpha' \rangle &=& - \frac{G^2_F |V_{cb}|^2}{6\pi^3}
\frac{m_{B'}^3}{\Gamma_T}\frac{ \int_1^{w_{\rm max}}
q^2\,(w^2-1)\,F_1(w)G_1(w) {\rm d }w}{1+2R_{L/T}} 
 ,
 \quad \langle \alpha^{\prime\prime} \rangle =
\frac{1-2R_{L/T}
 } {1+2R_{L/T}},  \qquad R_{L/T} = \frac{\Gamma_L}{\Gamma_T} \label{eq:rlt} 
\label{eq:averageasymmetry}
\end{eqnarray}

\subsection{Form factors}
\label{subsec:ff}
To obtain the form factors we have to evaluate the matrix elements
\begin{equation}
\left\langle B', r'\ \vec{p}^{\,\prime}\left|\,\overline \Psi^c(0)\gamma_\mu(I-\gamma_5)\Psi^b(0)
 \right| B, r\ \vec{p}
\right\rangle 
\end{equation}
which in our model are given by
\begin{equation}
\sqrt{\frac{E_{B}(\vec p\,)}{m_{B}}}\ \sqrt{\frac{E_{B'}(\vec
p\,')\,}{m_{B'}}}\
{}_{\stackrel{}{\stackrel{}{NR}}}\left\langle B', r'\ \vec{p}^{\,\prime}\left|\,\overline \Psi^c(0)\gamma_\mu(I-\gamma_5)\Psi^b(0)
 \right| B, r\ \vec{p}
\right\rangle_{NR}
\end{equation}
where the suffix ``$NR$'' denotes our nonrelativistic states and the
factors $\sqrt{E/m}$ take into account the different normalization.
We shall work in the initial baryon rest frame so that $\vec p=\vec
0,\,\vec p\,'=-\vec q$, and take $\vec q$ in the positive $z$
direction.  Furthermore we shall use the spectator
approximation. Having all this in mind we have in momentum space
\begin{eqnarray}
&&\hspace {-1.5cm}\sqrt{\frac{E_{B'}(-\vec q\,)}{m_{B'}}}\
{}_{\stackrel{}{\stackrel{}{NR}}}\left\langle B', r'\ -\vec{q}\left|\,\overline \Psi^c(0)\gamma_\mu(I-\gamma_5)\Psi^b(0)
 \right| B, r\ \vec{0}
\right\rangle_{NR}\nonumber\\
&&\hspace{1.cm}=\sqrt{2}\sqrt{\frac{E_{B'}(-\vec q\,)}{m_{B'}}}\ \sum_{s_1}\sum_{s_2} \left(\frac12\frac12 S_h
\bigg|s_1, s_2-s_1, s_2\right) \left(S_h\frac12\frac12
\bigg|s_2, r-s_2, r\right)\nonumber\\
&&\hspace{3.35cm}\times\left(\frac12\frac12 S'_h
\bigg|r'-r+s_1, s_2-s_1, r'-r+s_2\right) \left(S'_h\frac12\frac12
\bigg|r'-r+s_2, r-s_2, r'\right)\nonumber\\
&&\hspace{3.35cm}\times\int d\,^3q_1\, d\,^3q_2\, 
\left(\Phi^{B'}_{c\,h_2}(\vec q_1-\frac{m_{h_2}+m_q}{\overline M\,'}\,\vec q,
\,\vec{q}_2+\frac{m_{h_2}}{\overline M\,'}\,\vec q\,)\right)^*\ 
\Phi^B_{b\,h_2}(\vec{q}_1,\vec{q}_2)
\nonumber\\
&&\hspace{3.35cm}\times\sqrt{\frac{m_b}{E_b(\vec{q}_1)}}\sqrt{\frac{m_c}
{E_c(\vec{q}_1-\vec q\,)}}\ \bar{u}^c_{r'-r+s_1}(\vec{q}_1-\vec q\,)
\gamma_\mu(I-\gamma_5)\,u^b_{s_1}(\vec{q}_1)
\end{eqnarray}
where $\Phi^B_{b\,h_2}(\vec{q}_1,\vec{q}_2)$
($\Phi^{B'}_{c\,h_2}(\vec{q}_1,\vec{q}_2)$) is the Fourier transform
of the coordinate space wave function
$\Psi^B_{b\,h_2}(r_1,r_2,r_{12})$
($\Psi^{B'}_{c\,h_2}(r_1,r_2,r_{12})$) with $\vec q_1,\,\vec q_2$
being the conjugate momenta to the space variables $\vec r_1,\,\vec
r_2$. The factor of two comes from the fact that: i) for $bc-$baryon
decays, the charm quark resulting from the $b\to c$ transition could be
either the particle 1 or the particle 2 in the final $cc$ baryon,
while ii) for $bb-$baryon decays, there exist two equal contributions
resulting for the decay of each of the two bottom quarks of the
initial baryon.

The actual calculation is done in coordinate space where we have
\begin{eqnarray}
&&\hspace {-1.5cm}\sqrt{\frac{E_{B'}(-\vec q\,)}{m_{B'}}}\
{}_{\stackrel{}{\stackrel{}{NR}}}\left\langle B', r'\ -\vec{q}\left|\,\overline \Psi^c(0)\gamma_\mu(I-\gamma_5)\Psi^b(0)
 \right| B, r\ \vec{0}
\right\rangle_{NR}\nonumber\\
&&\hspace{1.cm}=\sqrt{2}\sqrt{\frac{E_{B'}(-\vec q\,)}{m_{B'}}}\ \sum_{s_1}\sum_{s_2} \left(\frac12\frac12 S_h
\bigg|s_1, s_2-s_1, s_2\right) \left(S_h\frac12\frac12
\bigg|s_2, r-s_2, r\right)\nonumber\\
&&\hspace{3.35cm}\times\left(\frac12\frac12 S'_h
\bigg|r'-r+s_1, s_2-s_1, r'-r+s_2\right) \left(S'_h\frac12\frac12
\bigg|r'-r+s_2, r-s_2, r'\right)\nonumber\\
&&\hspace{3.35cm}\times\int d\,^3r_1\, d\,^3r_2\, 
\Psi^{B'}_{c\,h_2}(r_1,r_2,r_{12})
\ e^{i\frac{m_{h_2}}{\overline M\,'}\,\vec
q\cdot\vec r_2}\,
e^{-i\frac{m_{h_2}+m_q}{\overline M\,'}\,\vec
q\cdot\vec r_1}
\nonumber\\
&&\hspace{3.35cm}\times\sqrt{\frac{m_b}{E_b(\vec{l}\ )}}\sqrt{\frac{m_c}
{E_c(\vec{l}-\vec q\,)}}\ \bar{u}^c_{r'-r+s_1}(\vec{l}-\vec q\,)
\gamma_\mu(I-\gamma_5)\,u^b_{s_1}(\vec{l}\ )\ \Psi^B_{b\,h_2}(r_1,r_2,r_{12})
\label{eq:coor}
\end{eqnarray}
where $\vec l=-i\stackrel{\rightarrow}{\nabla}_1$ represents an
internal momentum which is much smaller than the heavy quark masses
$m_b,\,m_c$. On the other hand $|\vec q\,|$ can be large\footnote{At
$q^2=0$ one has $|\vec q\,|=(m_B^2-m_{B'}^2) /2m_B$ which is $\approx
m_B/3$ for the transitions under study.}. Thus, to evaluate the above
expression we have made use of an expansion in $\vec l$, introduced in
Ref.~\cite{albertus05}, where second order terms in $\vec l$ are
neglected, while all orders in $|\vec q\,|$ are kept. For instance
$E_c(\vec l-\vec q\,)$ is approximated by $E_c(\vec l-\vec q\,)\approx
E_c(\vec q\,)\times (1-\vec l\cdot\vec q\,/E_c^2(\vec q\,))$ with
$E_c(\vec q\,)=\sqrt{m_c^2+\vec q\,^2}$.

The three vector and three axial form factors can be extracted from the set of
equations\footnote{Remember $\vec q$ is in the $z$ direction. Notice also that,
for $\vec p=\vec 0$, $w$ is just a function of $|\vec q\,|$.}
\begin{eqnarray}
\left\langle B',1/2\ -\vec{q}\left|\,\overline \Psi^c(0)\gamma_1\Psi^b(0)
 \right| B, -1/2\ \vec{0}
\right\rangle&=&\sqrt{\frac{E_{B'}(-\vec q\, )+m_{B'}}{2m_{B'}}}
\,\frac{|\vec q\,|}{E_{B'}(-\vec q\, )+m_{B'}}\ 
F_1(|\vec q\,|)\nonumber\\
\left\langle B', 1/2\ -\vec{q}\left|\,\overline \Psi^c(0)\gamma_3\Psi^b(0)
 \right| B, 1/2\ \vec{0}
\right\rangle&=&\sqrt{\frac{E_{B'}(-\vec q\, )+m_{B'}}{2m_{B'}}}\ 
|\vec q\,|\,\left(
\frac{F_1(|\vec q\,|)}{E_{B'}(-\vec q\, )+m_{B'}}+\frac{F_3(|\vec q\,|)}{m_{B'}}\right)\nonumber\\
\left\langle B', 1/2\ -\vec{q}\left|\,\overline \Psi^c(0)\gamma_0\Psi^b(0)
 \right| B, 1/2 \ \vec{0}
\right\rangle&=&\sqrt{\frac{E_{B'}(-\vec q\, )+m_{B'}}{2m_{B'}}}\left(
F_1(|\vec q\,|)+F_2(|\vec q\,|)+\frac{E_{B'}(-\vec q\, )}{m_{B'}}\,F_3(|\vec q\,|)\right)\nonumber\\
\label{eq:f123}
\end{eqnarray}\vspace{.25cm}
for the vector form factors and
\begin{eqnarray}
\left\langle B',1/2\  -\vec{q}\left|\,\overline \Psi^c(0)\gamma_1\gamma_5\Psi^b(0)
 \right| B, -1/2\ \vec{0}
\right\rangle&=&\sqrt{\frac{E_{B'}(-\vec q\, )+m_{B'}}{2m_{B'}}}
\ (-G_1(|\vec q\,|))\nonumber\\
\left\langle B', 1/2\ -\vec{q}\left|\,\overline \Psi^c(0)\gamma_3\gamma_5\Psi^b(0)
 \right| B, 1/2\ \vec{0}
\right\rangle&=&\sqrt{\frac{E_{B'}(-\vec q\, )+m_{B'}}{2m_{B'}}}\  \
\left(-G_1(|\vec q\,|)+\frac{|\vec q\,|^2\,G_3(|\vec q\,|)}{m_{B'}(E_{B'}(-\vec q\,
)+m_{B'})}\right)\nonumber\\
\left\langle B',1/2\  -\vec{q}\left|\,\overline \Psi^c(0)\gamma_0\gamma_5\Psi^b(0)
 \right| B,1/2\  \vec{0}
\right\rangle&=&\sqrt{\frac{E_{B'}(-\vec q\, )+m_{B'}}{2m_{B'}}}\,
\frac{|\vec q\,|}{E_{B'}(-\vec q\, )+m_{B'}}\bigg(
-G_1(|\vec q\,|)+G_2(|\vec q\,|)\nonumber\\
&&\hspace{5.75cm}+\frac{E_{B'}(-\vec q\, )}{m_{B'}}\,G_3(|\vec q\,|)\bigg)\nonumber\\
\label{eq:g123}
\end{eqnarray} 
for the axial ones. All the left hand side terms can be evaluated using Eq.(\ref{eq:coor}) with
the approximation mentioned above.

For each transition there are only two different coordinate space
integrals from which all different matrix elements can be
evaluated. Those integrals are
\begin{eqnarray}
{\cal I}^{B'B}(|\vec q\,|)&=&\int\, d\,^3r_1\, d\,^3r_2\, 
e^{i\frac{m_{h_2}}{\overline M\,'}\,\vec
q\cdot\vec r_2}\,
e^{-i\frac{m_{h_2}+m_q}{\overline M\,'}\,\vec
q\cdot\vec r_1}\ \left[\Psi^{B'}_{c\,h_2}(r_1,r_2,r_{12})\right]^*\ \Psi^{B}_{b\,h_2}(r_1,r_2,r_{12}) \nonumber\\
{\cal K}^{B'B}(|\vec q\,|)&=&\frac{1}{|\vec q\,|^2}\int\, d\,^3r_1\, d\,^3r_2\, 
e^{i\frac{m_{h_2}}{\overline M\,'}\,\vec
q\cdot\vec r_2}\,
e^{-i\frac{m_{h_2}+m_q}{\overline M\,'}\,\vec
q\cdot\vec r_1}\ \left[\Psi^{B'}_{c\,h_2}(r_1,r_2,r_{12})\right]^*\ 
\vec l\cdot\vec q\ \Psi^{B}_{b\,h_2}(r_1,r_2,r_{12})
\label{eq:integrales}
\end{eqnarray}
In appendix~\ref{app:fg123} we relate the form factors to the
 integrals ${\cal I}^{B'B}(|\vec q\,|)$ and ${\cal K}^{B'B}(|\vec
 q\,|)$ for the different $S_h,\,S'_h$ combinations, while in
 appendix~\ref{app:integrals} we give the actual expressions we use to
 evaluate those integrals.

\subsubsection{Current conservation}
In the limit $m_b=m_c$ and for $B'=B$ (and thus $S_h=S'_h$) vector
current conservation provides a relation among the vector $F_2$ and
$F_3$ form factors, namely
\begin{equation}
F_2(w)=F_3(w)
\label{eq:cc}
\end{equation}
On the other hand, if the two quarks in the current had the same
flavour, vector current conservation would fix the normalization of
the zeroth component of the vector current matrix element at $w=1$,
since this just counts the number of heavy quarks, two, in that case.
For quarks with distinct flavours, but still in the limit $m_b=m_c$,
the forward vector matrix element has an extra Clebsch-Gordan factor
of $1/\sqrt2$, and thus we have
\begin{equation}
F_1(1)+F_2(1)+F_3(1)=\sqrt{2}
\label{eq:baryonnumber}
\end{equation}
In this limiting situation  the integrals ${\cal I}^{BB}(|\vec q\,|)$
and ${\cal K}^{BB}(|\vec q\,|)$ are related by\footnote{One just has to
integrate by parts in the ${\cal K}^{BB}(|\vec q\,|)$ expression}%
\begin{equation}
{\cal K}^{BB}(|\vec q\,|)=\frac{m_{h_2}+m_q}{2\overline M}\
{\cal I}^{BB}(|\vec q\,|)
\end{equation}
Besides one has that ${\cal I}^{BB}(0)=1$. 

Using now the relations in Eq.(\ref{eq:appf123}) in
appendix~\ref{app:fg123} we see that our model satisfies the
constraint in Eq.(\ref{eq:baryonnumber}) exactly.  On the other hand
we have a small violation of vector  current conservation due to binding
effects.  For instance, and again using the
relations in Eq.(\ref{eq:appf123}), we obtain for $w=1$
\begin{equation}
F_2(1)=F_3(1)+\sqrt{2}(1-\frac{m_B}{\overline M})
\end{equation}
which shows that current conservation is violated by a term
proportional to the binding energy of the baryon divided by the sum of
the masses of its constituents.  This violation disappears in the infinite
heavy quark mass limit. This deficiency is avoided in most calculations  by 
the neglect of binding effects between the heavy diquark and the light quark\footnote{ In  Refs.~\cite{ebert04,
guo98} the currents
are constructed at the diquark level.  At
the baryon level, vector current is conserved because light quark-heavy diquark
binding effects are neglected. The calculation in
Ref.~\cite{sanchis95}  uses 
 the infinite heavy quark mass
limit thus canceling binding effects.}. Improvements on vector current
conservation would require at minimum the introduction of two--body
currents~\cite{buchmann94}, going thus beyond the spectator
approximation, that we have not considered in this analysis. 

Note also that, for transitions that do not conserve the spin 
of the heavy quark subsystem $S_h$
(i.e. $\Xi'^0_{bc} \to \Xi^+_{cc} l \bar\nu_l)$ we have in the 
$m_b=m_c$ limit and at zero recoil that
\begin{equation}
F_1(1)+F_2(1)+F_3(1)=0
\end{equation}
due to the orthogonality of the initial and final baryon wave--functions.

\subsection{Results}
\label{subsec:results}
\begin{figure}
\resizebox{12cm}{12cm}{\includegraphics{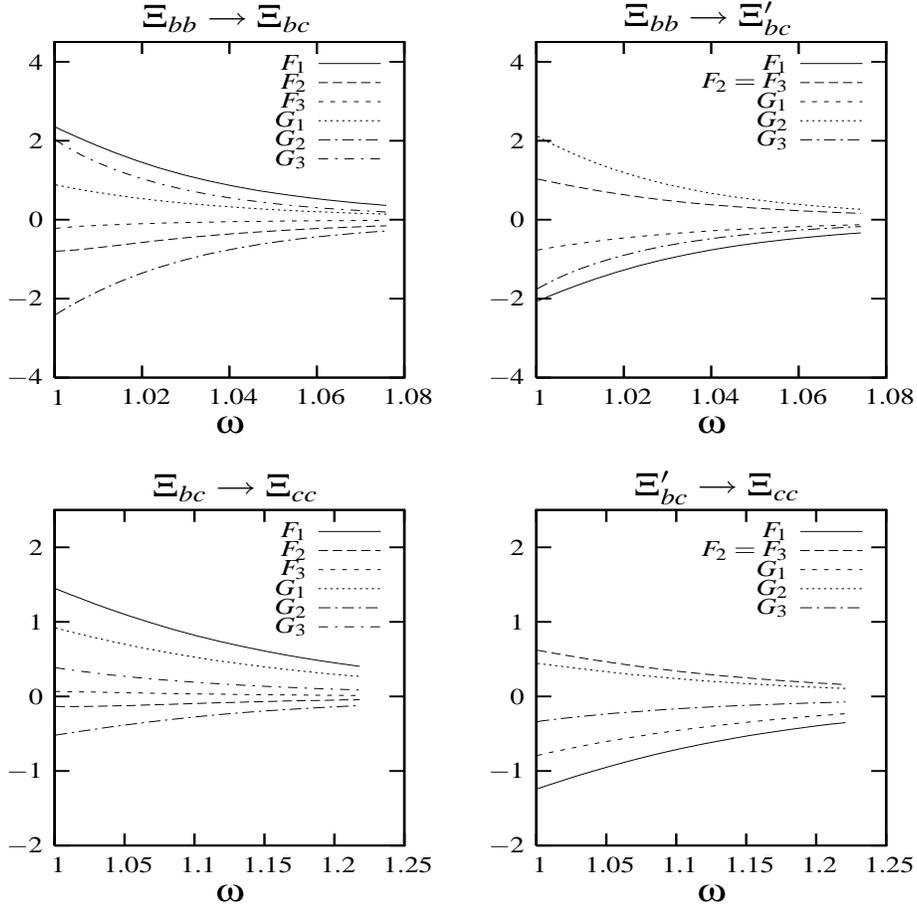}}
\caption{ Vector $F_1,\,F_2,\,F_3$ and axial $G_1,\,G_2,\,G_3$ form factors 
for doubly   $\Xi(J=1/2)$ baryons
decays evaluated with the AL1 potential. The results for $\Omega(J=1/2)$ baryons
(not shown) are very much the same.}
\label{fig:ffxi}
\end{figure}

In Fig.~\ref{fig:ffxi} we show the form factors
for $\Xi$ decays  evaluated with the AL1
potential. Variations when using a different potential are at the
level a few per cent at most. The results for doubly heavy $\Omega$
decays are almost identical to the corresponding ones for doubly heavy
$\Xi$ decays. The fact that we have two heavy quarks and that the
light one acts as a spectator makes the results almost independent of
the light quark mass.

\begin{figure}
\resizebox{13.cm}{12.cm}{\includegraphics{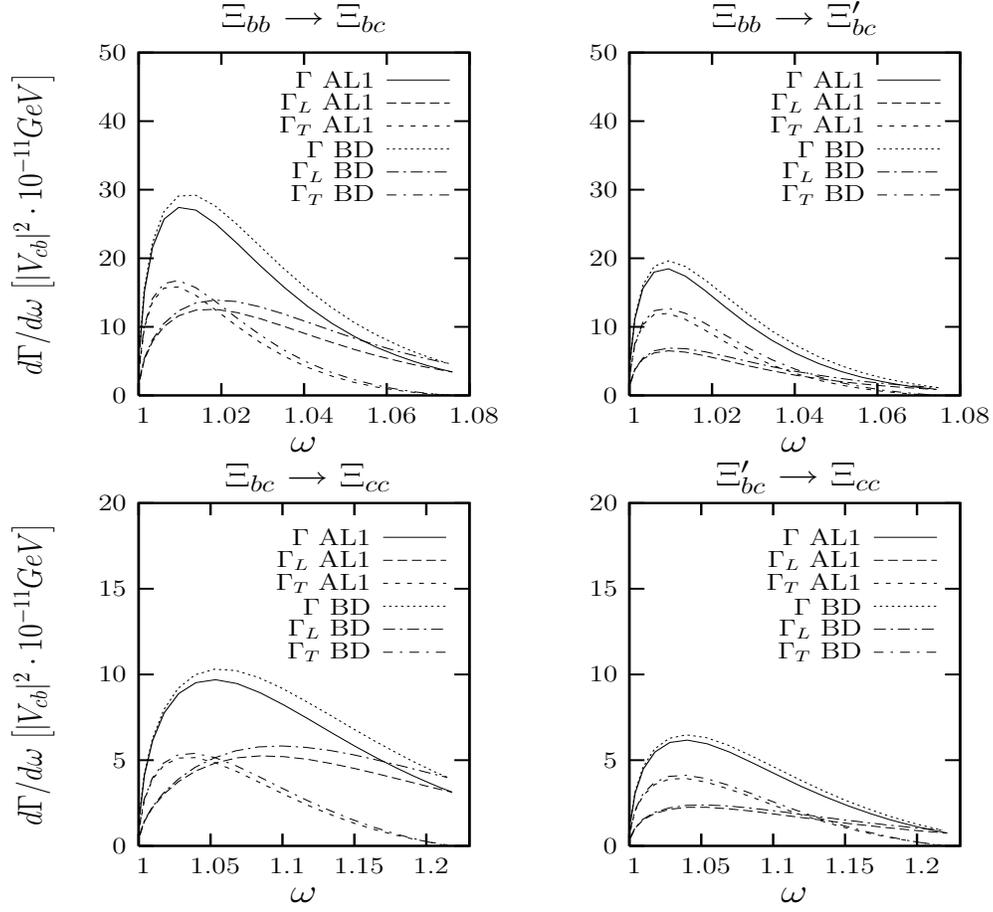}}\vspace{.5cm}\\
\caption{ $d\Gamma/dw$, $d\Gamma_L/dw$ and $d\Gamma_T/dw$ semileptonic
decay widths in units of $|V_{cb}|^2\cdot10^{-11}$\,GeV, for doubly
$\Xi(J=1/2)$ baryons decays. Solid line, long--dashed line and
short--dashed line: $d\Gamma/dw$, $d\Gamma_L/dw$ and $d\Gamma_T/dw$
evaluated with the AL1 potential; dotted line, long--dashed dotted
line and short--dashed dotted line: $d\Gamma/dw$, $d\Gamma_L/dw$ and
$d\Gamma_T/dw$ evaluated with the BD potential. The results for
$\Omega$ decay (not shown) are very similar.}
\label{fig:dgdwxi}
\end{figure}

In Fig.~\ref{fig:dgdwxi} we show now our results
for the differential $d\Gamma_T/dw$, $d\Gamma_L/dw$ and $d\Gamma/dw$
decay widths evaluated with the AL1 and BD potentials. The differences
between the results obtained with the two inter-quark interactions
could reach 30\% for some transitions and for some regions of $w$. As
a consequence of the apparent SU(3) symmetry in the form factors we
also find that the results for doubly heavy $\Xi$ and $\Omega$ decays (not
shown) are very close to each other. This apparent SU(3) symmetry goes over
to the integrated decay widths and asymmetry parameters.

In Table~\ref{tab:semi} we give our results for the semileptonic decay
width (transverse $\Gamma_T$, longitudinal $\Gamma_L$ and total
$\Gamma$) for the different processes under study. Our central values
have been evaluated using the AL1 potential while the errors show the
variations when changing the interaction. The biggest variations
appear for the BD potential for which one obtains results which are
larger by $7\sim12\% 
$. In Table~\ref{tab:gamma} we compare our
results to the ones calculated in different models. For that purpose
we need a value for $|V_{cb}|$ for which we take $|V_{cb}|=0.0413$.
Our results are in reasonable agreement with the ones in
Ref.~\cite{ebert04} where they use a relativistic quark model
evaluated in the quark-diquark approximation\footnote{ 
Note we have divided by a factor 2 the results originally published in Ref.~\cite{ebert04}
as there, and  as we initially did, the authors forgot a symmetry factor $1/\sqrt2$\ for the case of diquarks with two
equal quarks~\cite{ebert08}.}.
The result for 
$\Gamma(\Xi_{bc}\to\Xi_{cc})$ obtained in  Ref.~\cite{sanchis95}
using HQET is a factor of 4 larger than ours\footnote{Note we have multiplied by a factor 2 the result originally published in Ref.~\cite{sanchis95}
as the author overlooked a normalization factor $\sqrt2$~\cite{sanchis08}.}.
We can not directly  compare our result for the latter transition with the one published in  Ref.~\cite{faessler01}, the
reason being that in this  reference the $\Xi_{bc}$ state is defined differently, with the $c$ and the
light quark coupled to total spin 1. Had they defined the state with the $bc$ pair coupled to spin 1 they
would had obtained  roughly a factor of 4 larger result~\cite{gutsche08}, in reasonable agreement with our calculation or the
one by Ebert {\it et al.}\footnote{Note the actual physical states would be an admixture of what we call 
$\Xi_{bc}\,\Xi'_{bc}$, and similarly in the $\Omega$ case. A measurement of the decay width could give
information on the admixtures.}. Finally in
Ref.~\cite{guo98}, where they use the Bethe--Salpeter equation applied
to a quark-diquark system, they obtain much larger results for all
transitions.
\begin{table}
\begin{tabular}{lccc}
  &$\Gamma_T$ & $\Gamma_L$& $\Gamma$ \\ \hline\\
$\Xi_{bb}\to\Xi_{bc}\,l\bar\nu_l$\hspace{1cm}  & $0.49^{+0.5}_{-0.01}$ &
$0.64^{+0.10}_{-0.02} $
&$ 1.13^{+0.15}_{-0.03}$\\ \\
$\Xi_{bc}\to\Xi_{cc}\,l\bar\nu_l$  & $0.58^{+0.04}_{-0.01}$ &$ 0.93^{+0.11}_{-0.01}$
&$ 1.51^{+0.15}_{-0.02} $ \\ \\
$\Xi_{bb}\to\Xi_{bc}'\,l\bar\nu_l$ & $0.37^{+0.04}_{-0.01}$ & $0.26^{+0.04}_{-0.01} $
& $0.62 ^{+0.08}_{-0.02}$ \\ \\
$\Xi_{bc}'\to\Xi_{cc}\,l\bar\nu_l$ & $0.45^{+0.03}_{-0.01}$ & $0.35^{+0.03}_{-0.01} $
 & $0.80^{+0.06}_{-0.02}$ \\ \hline
\end{tabular}\hspace{1cm}
\begin{tabular}{lccc}
  &$\Gamma_T$ & $\Gamma_L$& $\Gamma$ \\ \hline\\
$\Omega_{bb}\to\Omega_{bc}\,l\bar\nu_l$\hspace{1cm}  & $0.53^{+0.04}_{-0.01}$&
$0.73^{+0.08}_{-0.01}$ & $1.26^{+0.12}_{-0.01}$\\ \\
$\Omega_{bc}\to\Omega_{cc}\,l\bar\nu_l$  &$ 0.58^{+0.03}$ &$0.94^{+0.09}$
&$1.52^{+0.12}$\\ \\
$\Omega_{bb}\to\Omega_{bc}'\,l\bar\nu_l$  & $0.40^{+0.04}$ & $0.29^{+0.04} $
&$0.68^{+0.08}$\\ \\
$\Omega_{bc}'\to\Omega_{cc}\,l\bar\nu_l$  & $0.45^{+0.03}$ &
$0.35^{+0.03}$&$0.80^{+0.05} $ \\\hline
\end{tabular}
\caption{Semileptonic decay widths in units of
$|V_{cb}|^2\cdot10^{-11}$\,GeV. $\Gamma_T$ and $\Gamma_L$ stand for the
transverse and longitudinal contributions to the width $\Gamma$.
The central values have been obtained with the AL1 potential.
$l$ stands for a light charged lepton, $l=e,\,\mu$}
\label{tab:semi}
\end{table}
\begin{table}
\begin{tabular}{lccccc}
  &This work &\cite{ebert04}$^\dag$&\cite{faessler01}$^\ddag$&\cite{guo98}&\cite{sanchis95}$^\S$\\ \hline\\
$\Gamma(\Xi_{bb}\to\Xi_{bc}\,l\bar\nu_l)$\hspace{1cm}  &$ 1.92^{+0.25}_{-0.05}$&
1.63&&28.5&\\ \\
$\Gamma(\Xi_{bc}\to\Xi_{cc}\,l\bar\nu_l)$  &$ 2.57^{+0.26}_{-0.03} $ & 2.30&0.79&8.93&8.0\\ \\
$\Gamma(\Xi_{bb}\to\Xi_{bc}'\,l\bar\nu_l)$ &  $1.06 ^{+0.13}_{-0.03}$ &0.82&&4.28&\\ \\
$\Gamma(\Xi_{bc}'\to\Xi_{cc}\,l\bar\nu_l)$ &  $1.36^{+0.10}_{-0.03}$ & 0.88&&7.76&\\ \hline
\end{tabular}\hspace{1cm}
\begin{tabular}{lccc}
  &This work &\cite{ebert04}$^\dag$\\ \hline\\
$\Gamma(\Omega_{bb}\to\Omega_{bc}\,l\bar\nu_l)$\hspace{1cm}  & $2.14^{+0.20}_{-0.02}$& 1.70\\ \\
$\Gamma(\Omega_{bc}\to\Omega_{cc}\,l\bar\nu_l)$ &$2.59^{+0.20}$&2.48\\ \\
$\Gamma(\Omega_{bb}\to\Omega_{bc}'\,l\bar\nu_l)$ &$1.16^{+0.13}$& 0.83\\ \\
  $\Gamma(\Omega_{bc}'\to\Omega_{cc}\,l\bar\nu_l)$ &$1.36^{+0.09} $ &0.95\\\hline
\end{tabular}
\caption{Semileptonic decay widths in units of $10^{-14}$\,GeV. We have
used a value $|V_{cb}|=0.0413$. $l$ stands for a light charged lepton, $l=e,\,\mu$. For results
with $\dag,\,\ddag$ and $\S$ see text for details.}
\label{tab:gamma}
\end{table}%

In Table~\ref{tab:asimetrias} we compile our results for the average
angular asymmetries $\alpha'$ and $\alpha''$, as well as the $R_{L/T}$
ratio, introduced in Eq.(\ref{eq:averageasymmetry}). The central
values have been obtained with the AL1 potential. Being all quantities
ratios the variation when changing the inter-quark interaction are in
most cases small.
\begin{table}
\begin{tabular}{lccc}
& $\left<\alpha'\right>$ &
  $\left<\alpha''\right>$ &$R_{L/T}$\\ \hline \\
$\Xi_{bb}\to\Xi_{bc}\,l\bar\nu_l$\hspace{1cm}  &$-0.13^{+0.01}$&$-0.45_{-0.02}$&
$1.33^{+0.06}_{-0.01}$ \\\\

$\Xi_{bc}\to\Xi_{cc}\,l\bar\nu_l$  &$-0.12^{+0.01}$&$-0.53_{-0.01}$&
$1.62^{+0.07}$ \\\\

 $\Xi_{bb}\to\Xi_{bc}'\,l\bar\nu_l$   & $-0.19$ & $-0.17_{-0.01}$ &
$0.71^{+0.01}_{-0.01}$ \\ \\

 $\Xi_{bc}'\to\Xi_{cc}\,l\bar\nu_l$  & $-0.19$ & $-0.23_{-0.01}$ &
$0.79^{+0.02}$  \\ \hline
\end{tabular}\hspace{1cm}
\begin{tabular}{lccc}
& $\left<\alpha'\right>$ &
  $\left<\alpha''\right>$ &$R_{L/T}$\\ \hline \\
$\Omega_{bb}\to\Omega_{bc}\,l\bar\nu_l$ \hspace{1cm}
 &$ -0.13^{+0.01}$ & $-0.47_{-0.01}$&$1.37^{+0.06}$ \\ \\
%

%
 $\Omega_{bc}\to\Omega_{cc}\,l\bar\nu_l$  
 & $-0.12^{+0.01}$ &$-0.53_{-0.01}$ &$1.63^{+0.06}$
\\ \\

 $\Omega_{bb}\to\Omega_{bc}'\,l\bar\nu_l$ 
 & $-0.19_{-0.01}$ & $-0.18_{-0.01}$&$0.72^{+0.02}$
\\ \\
%

 $\Omega_{bc}'\to\Omega_{cc}\,l\bar\nu_l$ 
  & $-0.19$ & $-0.23_{-0.01}$ & $0.79^{+0.02}$\\\hline

\end{tabular}
\caption{Averaged values of the asymmetry parameters $\alpha'$ and $\alpha''$
evaluated as indicated in 
Eq.(\ref{eq:averageasymmetry}). We also show the ratio $R_{L/T}=\Gamma_L/\Gamma_T$.
The central values have been obtained with the AL1 potential. 
$l$ stands for a light charged lepton, $l=e,\,\mu$}
\label{tab:asimetrias}
\end{table}

\section{Summary}
\label{sec:summary}

We have evaluated static properties and semileptonic decays for the
ground state of doubly heavy $\Xi$ and $\Omega$ baryons. The
calculations have been done in the framework of a nonrelativistic
quark model with the use of five different inter-quark
interactions. The use of different quark-quark potentials allows us to
obtain an estimation of the theoretical uncertainties related to the
quark-quark interaction.  In order to
build our wave functions we have made use of the constraints imposed
by the infinite heavy quark mass limit. In this limit the spin--spin
interactions vanish and the total spin of the two heavy quarks is well
defined. With this approximation we have used a simple variational
approach, with Jastrow type orbital wave functions, to solve the
involved three-body problem.
 
 Among the static properties, our results for the masses are in very
 good agreement with previous results obtained with the same
 inter-quark interactions but within a more complicated Faddeev
 approach~\cite{Si96}. In some cases we even get lower, and thus
 better, masses\footnote{For a given Hamiltonian a variational 
 mass is an upper bound of
 the true ground state mass.}. We have calculated mass densities and charge
 densities (charge form factors) finding that the corresponding mean
 square radii are again in good agreement with the Faddeev calculation
 of Ref.~\cite{Si96}.  We have also evaluated magnetic moments. Being
 the total orbital angular momentum of the baryon $L=0$, the magnetic
 moments come from the spin contributions alone.  With the exception
 of $\Xi_{bc}^0$, $\Xi_{bc}^+$ and $\Omega_{bc}^0$ we agree perfectly
 with the Faddeev calculation in Ref.~\cite{Si96}. For the magnetic
 moments of $\Xi_{bc}^0$, $\Xi_{bc}^+$ and $\Omega_{bc}^0$ the
 discrepancies between the two calculation are very large. The origin
 might be attributed to the presence of a non--negligible $S_h=0$ component in
 the wave functions of Ref.~\cite{Si96}. In our case we have $S_h=1$
 which we think is a good approximation based on the infinite heavy
 quark mass limit. This assertion seems to be corroborated by the
 results obtained in the relativistic calculation of
 Ref.~\cite{faessler06}, at least for the $\Xi_{bc}^0$ and
 $\Omega_{bc}^0$ cases.

We have used our  wave functions to study the semileptonic decay
of doubly $\Xi(J=1/2)$ and $\Omega(J=1/2)$ baryons.  We have worked in
the spectator approximation with one--body currents alone.  In the
$m_b=m_c$ case and for $B=B'$ baryons we have checked that our model
satisfies baryon number conservation. On the other hand we have a
small vector current violation by an amount given by the binding
energy over the mass of the baryon, violation that disappears in the infinite
heavy quark mass limit. Small vector current violations due to binding effects 
seems to be present in most calculations to date.
 With this model we have evaluated
form factors, asymmetry parameters, differential decay widths and
total decay widths. Our results for the latter are in reasonable
agreement with the ones obtained in Ref.~\cite{ebert04} using a
relativistic quark model in the quark--diquark approximation, while
they are much smaller than the ones obtained in Ref.~\cite{guo98} by
means of the Bethe--Salpeter equation applied to a quark-diquark
system.

For the weak form factors the results exhibit an apparent SU(3)
symmetry when going from $\Xi$ to $\Omega$ baryons. This is due to the
fact that we have two heavy quarks and the light one acts as a
spectator in the weak transition. This apparent symmetry appears also
in the decay widths and asymmetry parameters. On the other hand SU(3)
violating effects are clearly visible in some static quantities like
the charge form factors and radii, 
and the magnetic moments, that depend strongly on the light quark
charge and/or mass. \\

\begin{acknowledgments}
 This research was supported by DGI and FEDER funds, under contracts
FIS2005-00810, BFM2003-00856 and FPA2004-05616, by Junta de
Andaluc\'\i a and Junta de Castilla y Le\'on under contracts FQM0225
and SA104/04, and it is part of the EU integrated infrastructure
initiative Hadron Physics Project under contract number
RII3-CT-2004-506078.  J. M. V.-V. acknowledges an E.P.I.F. contract  with the
University of Salamanca.

\end{acknowledgments}

\appendix
\section{Infinite heavy quark mass limit of our variational wave functions}
\label{app:ihqml}
In the infinite heavy quark mass limit the wave function of the system should
look like the one for a ``meson'' composed of a light quark and a heavy
diquark. The two heavy quarks bind into a $\bar 3$ color source diquark 
that to the light degrees of
freedom appears to be pointlike~\cite{white91}. In our model the pointlike
nature of the heavy diquark  comes about
through the one-gluon exchange Coulomb potential which binds the two heavy
quarks into a distance\footnote{The relation can  only be approximate due to 
confinement and the interaction with the light
quark.}
\begin{eqnarray}
r_{h_1h_2}\propto\frac{1}{\mu_{h_1h_2}};\ 
\mu_{h_1h_2}=\frac{m_{h_1}m_{h_2}}{m_{h_1}+m_{h_2}}
\end{eqnarray} 
that tends to zero if both quark masses go to infinity.

\begin{figure}
\resizebox{12.cm}{!}{\includegraphics{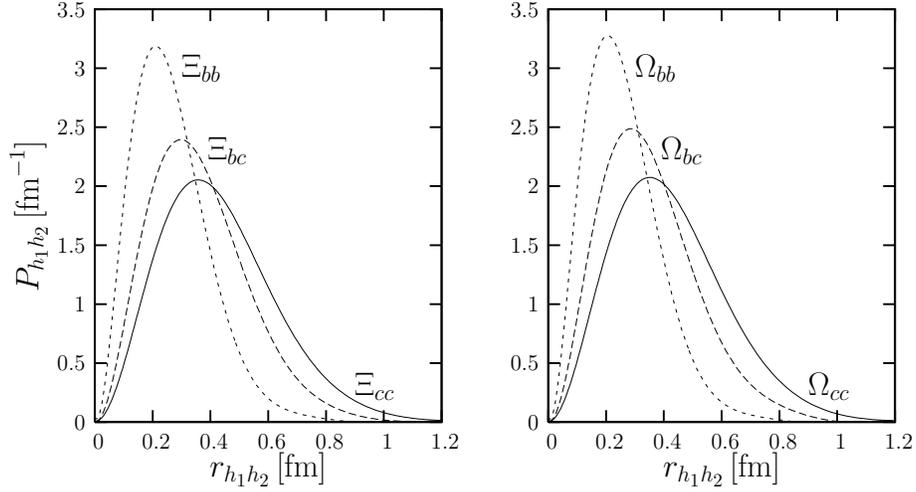}}
\caption{ $P_{h_1h_2}$ probabilities (see Eq.(\ref{eq:ph1h2})) for the
$\Xi_{cc},\,\Xi_{bc},\, \Xi_{bb}$ and 
$\Omega_{cc},\,\Omega_{bc},\, \Omega_{bb}$ baryons evaluated with the AL1
potential.  }
\label{fig:p12}
\end{figure}

For our wave functions we can define the probability distribution
$P_{h1h2}$
for the two heavy quarks
to be found at a distance $r_{h_1h_2}$
\begin{equation}
P_{h1h2}\,(r_{h_1h_2}) =\int d^3r_1\int d^3r_2\ \delta(r_{12}-r_{h_1h_2})
\left|\Psi^B_{h_1h_2}(r_2,r_2,r_{12})
\right|^2
\label{eq:ph1h2}
\end{equation}
In Fig.~\ref{fig:p12}, we show the $P_{h1h2}$ probability distributions
 for the $\Xi_{cc},\,\Xi_{bc},\, \Xi_{bb}$ and 
$\Omega_{cc},\,\Omega_{bc},\, \Omega_{bb}$ baryons evaluated 
for the AL1 potential.
We see how the maximum moves, as expected, to lower $r_{h_1h_2}$ values as the quark masses
increase.

On the other hand the relative wave function for the light quark with respect to the heavy
quark subsystem should tend to the one of a light quark relative to a pointlike
diquark. This limit is not evident in the coordinates we work as one
has  first to separate the diquark internal orbital wave function from the 
total one. 
To see that this limiting situation is in fact reached within our variational
ansatz let us do the following: introduce the new set of coordinates
\begin{eqnarray}
\vec R &=&\frac{m_{h_1}\vec x_{h_1}+m_{h_2}\vec x_{h_2}+m_{q}\vec x_{q}}
{m_{h_1}+m_{h_2}+m_{q}}\nonumber\\
\vec r_{12}&=&\vec x_{h_1}-\vec x_{h_2}\nonumber\\
\vec r_q&=&\frac{m_{h_1}\vec x_{h_1}+m_{h_2}\vec x_{h_2}}{m_{h_1}+m_{h_2}}-\vec
x_q=\frac{m_{h_1}\vec r_{1}+m_{h_2}\vec r_{2}}{m_{h_1}+m_{h_2}}
\end{eqnarray} 
in terms of which the Hamiltonian now reads
\begin{eqnarray}
H&=&
-\frac{\stackrel{\rightarrow}{\nabla}
\stackrel{}{^2}_{\hspace{-.1cm}\vec{R}}}{2 \overline M} +
H^{\rm int} \nonumber\\
 H^{\rm
int}&=&\overline M+H_{h1h_2}+H_{qh_1h_2}
\end{eqnarray}
where
\begin{eqnarray}
H_{h1h_2}&=&-\frac{\stackrel{\rightarrow}{\nabla}_{12}^2}{2\mu_{h_1h_2}}
+V_{h_1h_2}(\vec r_{12},\, spin)\nonumber\\
 H_{qh_1h_2}&=&-\frac12\left(\frac{1}{m_{h_1}+m_{h_2}}+\frac{1}{m_q}\right)
 \stackrel{\rightarrow}{\nabla}_{q}^2
+V_{h_1q}(\vec r_q+\frac{m_{h_2}}{m_{h_1}+m_{h_2}}\vec r_{12},\,
spin)\nonumber\\
&&\hspace{4.55cm}+V_{h_2q}(\vec r_q-\frac{m_{h_1}}{m_{h_1}+m_{h_2}}\vec r_{12},\, spin)
\end{eqnarray}
with $\stackrel{\rightarrow}{\nabla}_{12}=\partial /\partial_{\vec r_{12}}$ and
$\stackrel{\rightarrow}{\nabla}_{q}=\partial /\partial_{\vec r_q}$.
Defining now
\begin{eqnarray}
 H^0_{qh_1h_2}&=&-\frac12\left(\frac{1}{m_{h_1}+m_{h_2}}+\frac{1}{m_q}\right)
 \stackrel{\rightarrow}{\nabla}_{q}^2
+V_{h_1q}(\vec r_q,\,
spin)+V_{h_2q}(\vec r_q,\, spin)
\end{eqnarray}
one would have
\begin{eqnarray}
H^{\mathrm{int}}=H_{h_1h_2}+H^0_{qh_1h_2}+(H_{qh_1h_2}-H^0_{qh_1h_2})
\end{eqnarray}
$H_{h_1h_2}$ is the  Hamiltonian for the relative
 movement of the two heavy
quarks while $H^0_{qh_1h_2}$ is the Hamiltonian  for the
relative movement of the light quark with respect to a pointlike  heavy diquark
  where the two heavy quarks are  located in their center of mass. 
Both movements are coupled through the term
$(H_{qh_1h_2}-H^0_{qh_1h_2})$.
If the  heavy quark masses get arbitrarily  large  the average $r_{12}$ value
tends to zero and thus one can neglect the effect of
$H_{qh_1h_2}-H^0_{qh_1h_2}$. In that limiting situation the light and heavy quark degrees
of freedom decouple completely and the internal Hamiltonian reduces, as it should, to the sum
of the Hamiltonians $H_{h_1h_2}$, corresponding  to the relative movement of the 
two heavy quarks,  and $H^0_{qh_1h_2}$, corresponding to the  relative movement of the light quark
with respect to the pointlike heavy diquark subsystem. 

As to the wave function it should reduce in that limit to the product 
$\Phi_{h_1h_2}(r_{12})\cdot \Phi^0_{qh_1h_2}(r_q)$, being $\Phi_{h_1h_2}(r_{12}),\, \Phi^0_{qh_1h_2}(r_q)$ the ground state wave
functions for $H_{h_1h_2},\,H^0_{qh_1h_2}$ respectively. To check that we reach
that limiting situation we have evaluated
the projection ${\cal P}$ of our variational wave functions  onto
  $\Phi_{h_1h_2}(r_{12})\cdot \Phi^0_{qh_1h_2}(r_q)$ evaluated with the actual
  heavy quark masses.
  This projection is
given by\footnote{Note $r_q$ can be expressed in terms of 
$r_1,r_2,r_{12}$ and that $d^3r_1d^3r_2=d^3r_{12}d^3r_q$}
\begin{equation}
{\cal P}=\int d^3r_1 \int d^3r_2 \left(\Psi^B_{h_1h_2}(r_1,r_2,r_{12}\right)^*
\Phi_{h_1h_2}(r_{12})\ \Phi^0_{qh_1h_2}(r_q)
\label{eq:project}
\end{equation}
and the values for $|{\cal P}|^2$ that we obtain for the $\Xi_{cc},\,\Xi_{bc},\, \Xi_{bb}$ and 
$\Omega_{cc},\,\Omega_{bc},\, \Omega_{bb}$ baryons using the AL1 potential are given in
Table~\ref{tab:project}. We see how $|{\cal P}|^2$ increases with increasing heavy
quark masses indicating that for very high heavy quark masses the total wave
function tends to the one of a light quark relative to a pointlike
diquark times the diquark internal wave function. 
\begin{table}[h!!!]
\begin{tabular}{lccc|ccc}
&$\Xi_{cc}$&\hspace{.3cm}$\Xi_{bc}$&\hspace{.3cm}$\Xi_{bb}$
&\hspace{.3cm}$\Omega_{cc}$&\hspace{.3cm}$\Omega_{bc}$&\hspace{.3cm}$\Omega_{bb}$\\
\hline
$|{\cal P}|^2$ &0.974&0.975&0.991&0.959&0.966&0.984
\end{tabular}
\caption{Absolute value square of the ${\cal P}$ projection coefficient  defined in
Eq.~(\ref{eq:project})}
\label{tab:project}
\end{table}%
 In summary, our variational wave functions respect the infinite heavy quark
 mass limit.

We note by passing that the product $\Phi_{h_1h_2}(r_{12})\cdot \Phi^0_{qh_1h_2}(r_q)$ corrected by
a correlation function  in the variable $\vec r_{12}\cdot \vec r_q$ would have
 been a good   variational orbital wave function. 
For instance,  a calculation using the AL1 potential  and  an orbital wave 
function given just by
 that product $\Phi_{h_1h_2}(r_{12})\cdot \Phi^0_{qh_1h_2}(r_q)$  gives for  the
 expectation value of $H^{\mathrm{int}}$ for the $\Xi_{cc},\,\Xi_{bc},\,\Xi_{bb}$
 baryons
 \begin{eqnarray}
 \left.\langle H^{\mathrm{int}}\rangle\right|^{AL1}_{\Xi_{cc}}=3640\,\mathrm{MeV}; \
 \left.\langle H^{\mathrm{int}}\rangle\right|^{AL1}_{\Xi_{bc}}=6943\,\mathrm{MeV};\
 \left.\langle H^{\mathrm{int}}\rangle\right|^{AL1}_{\Xi_{bb}}=10198\,\mathrm{MeV}
 \end{eqnarray}
 which are respectively 28\,MeV, 24\,MeV, and 1\,MeV larger, and therefore worse (see footnote
 13), than our best values in
 Table~\ref{tab:xiomegamasses}. The correlation function would clearly be 
 needed,
 being its role more important for the ``$cc$'' and ``$bc$'' systems.
 The improvement in going from a ``$cc$''  to a ``$bc$ system'' is not as good as
 one would naively expect, the reason being that,
 considering $\vec r_{12}$ to be a small quantity,
$H_{qh_1h_2}-H^0_{qh_1h_2}$ is first order in $\vec r_{12}$ for the ``$bc$''
system while, for symmetry reasons,  it is necessarily of second order
 for the ``$cc$'' and ``$bb$'' ones. In any case one sees how the expectation 
 values obtained with this
 simple ansatz get closer to our best values in Table~\ref{tab:xiomegamasses}
 as the heavy quark masses increase.

\section{Form factors in terms of the ${\cal I}^{B'B}(|\vec q\,|)$ 
and ${\cal K}^{B'B}(|\vec q\,|)$
integrals}
\label{app:fg123}
In this appendix we relate the vector  $F_1,\,F_2,\,F_3$
and axial 
$G_1,\,G_2,\,G_3$
form factors, that we evaluate in the center of mass of the decaying baryon, to the integrals ${\cal I}^{B'B}(|\vec q\,|),
\,{\cal K}^{B'B}(|\vec q\,|)$ defined in Eq.(\ref{eq:integrales}).  To simplify the expressions it is convenient
to introduce
\begin{eqnarray}
\widehat F_j(|\vec q\,|)&=&\sqrt{\frac{E_{B'}(-\vec q\, )+m_{B'}}{2E_{B'}(-\vec q\, )}}
\,\sqrt{\frac{2E_{c}(\vec q\, )}{E_{c}(\vec q\, )+m_{c}}}\ F_j(|\vec q\,|)\ \ ,\ j=1,2,3\nonumber\\
\widehat G_j(|\vec q\,|)&=&\sqrt{\frac{E_{B'}(-\vec q\, )+m_{B'}}{2E_{B'}(-\vec q\, )}}
\,\sqrt{\frac{2E_{c}(\vec q\, )}{E_{c}(\vec q\, )+m_{c}}}\ G_j(|\vec q\,|)\ \ ,\ j=1,2,3\nonumber\\
\end{eqnarray}

\begin{itemize}
\item Cases $S_h=1,\,S'_h=0$ or $S_h=0,\,S'_h=1$:
\begin{eqnarray}
\frac{\widehat F_1(|\vec q\,|)}{E_{B'}(-\vec q\, )+m_{B'}}\ 
&=&-\sqrt\frac{2}{3}\,
\left(\frac{{\cal I}^{B'B}(|\vec q\,|)}{E_{c}(\vec q\, )+m_c}
-\frac{{\cal K}^{B'B}(|\vec q\,|)}{2}
\left(\frac{m_c}{E^2_{c}(\vec q\, )}-\frac{1}{m_b}
\right)
\right)
\nonumber\\
\frac{\widehat F_1(|\vec q\,|)}{E_{B'}(-\vec q\, )+m_{B'}}+\frac{\widehat F_3(|\vec q\,|)}{m_{B'}}&=&0\nonumber\\
\widehat F_1(|\vec q\,|)+\widehat F_2(|\vec q\,|)+\frac{E_{B'}(-\vec q\, )}{m_{B'}}\,\widehat F_3(|\vec q\,|)&=&0\nonumber\\
\label{eq:appf123s10}
\end{eqnarray} 
where $E_c(\vec q\,)=\sqrt{m_c^2+\vec q^{\,2}}$. From the above expressions
we have that $F_2=F_3$.

\begin{eqnarray}
-\widehat G_1(|\vec q\,|)&=&\sqrt\frac{2}{3}\,
\left(\,{\cal I}^{B'B}(|\vec q\,|)+\frac{|\vec q\,|^2\,{\cal K}^{B'B}(|\vec q\,|)}{2\left(E_{c}(\vec q\, )+m_c\right)}
\left(\frac{m_c}{E^2_{c}(\vec q\, )}+\frac{1}{m_b}
\right)
\right)\nonumber\\
\left(-\widehat G_1(|\vec q\,|)+\frac{|\vec q\,|^2\,\widehat G_3(|\vec q\,|)}{m_{B'}(E_{B'}(-\vec q\,
)+m_{B'})}\right)&=&\sqrt\frac{2}{3}\,
\left(\,{\cal I}^{B'B}(|\vec q\,|)+\frac{|\vec q\,|^2\,{\cal K}^{B'B}(|\vec q\,|)}{2\left(E_{c}(\vec q\, )+m_c\right)}
\left(\frac{m_c}{E^2_{c}(\vec q\, )}-\frac{1}{m_b}
\right)
\right)\nonumber\\
\frac{-\widehat G_1(|\vec q\,|)+\widehat G_2(|\vec q\,|)+\frac{E_{B'}(-\vec q\, )}{m_{B'}}\,
\widehat G_3(|\vec q\,|)}{E_{B'}(-\vec q\, )+m_{B'}}&=&\sqrt\frac{2}{3}\,
\left(\,\frac{{\cal I}^{B'B}(|\vec q\,|)}{E_{c}(\vec q\, )+m_c}
-\frac{{\cal K}^{B'B}(|\vec q\,|)}{2}
\left(\frac{m_c}{E^2_{c}(\vec q\, )}+\frac{1}{m_b}
\right)
\right)\nonumber\\ 
\end{eqnarray}

\item Case $S_h=1,\,S'_h=1$

\begin{eqnarray}
\frac{\widehat F_1(|\vec q\,|)}{E_{B'}(-\vec q\, )+m_{B'}}\ 
&=&\frac{4}{3\sqrt2}\,
\left(\frac{{\cal I}^{B'B}(|\vec q\,|)}{E_{c}(\vec q\, )+m_c}
-\frac{{\cal K}^{B'B}(|\vec q\,|)}{2}
\left(\frac{m_c}{E^2_{c}(\vec q\, )}-\frac{1}{m_b}
\right)
\right)
\nonumber\\
\frac{\widehat F_1(|\vec q\,|)}{E_{B'}(-\vec q\, )+m_{B'}}+
\frac{\widehat F_3(|\vec q\,|)}{m_{B'}}&=&
\sqrt2\,
\left(\frac{{\cal I}^{B'B}(|\vec q\,|)}{E_{c}(\vec q\, )+m_c}
-\frac{{\cal K}^{B'B}(|\vec q\,|)}{2}
\left(\frac{m_c}{E^2_{c}(\vec q\, )}+\frac{1}{m_b}
\right)
\right)
\nonumber\\
\widehat F_1(|\vec q\,|)+\widehat F_2(|\vec q\,|)+\frac{E_{B'}(-\vec q\, )}{m_{B'}}\,\widehat F_3(|\vec q\,|)&=&
\sqrt2\,
\left(\,{\cal I}^{B'B}(|\vec q\,|)
+\frac{|\vec q\,|^2\,{\cal K}^{B'B}(|\vec q\,|)}
{2\left(E_{c}(\vec q\, )+m_c\right)}
\left(\frac{m_c}{E^2_{c}(\vec q\, )}-\frac{1}{m_b}
\right)
\right)
\nonumber\\
\label{eq:appf123}
\end{eqnarray} 

\begin{eqnarray}
\widehat G_1(|\vec q\,|)&=&\frac{4}{3\sqrt2}\,
\left(\,{\cal I}^{B'B}(|\vec q\,|)+\frac{|\vec q\,|^2\,{\cal K}^{B'B}(|\vec q\,|)}
{2\left(E_{c}(\vec q\, )+m_c\right)}
\left(\frac{m_c}{E^2_{c}(\vec q\, )}+\frac{1}{m_b}
\right)
\right)\nonumber\\
\left(-\widehat G_1(|\vec q\,|)+\frac{|\vec q\,|^2\,\widehat G_3(|\vec q\,|)}{m_{B'}(E_{B'}(-\vec q\,
)+m_{B'})}\right)&=&-\frac{4}{3\sqrt2}\,
\left(\,{\cal I}^{B'B}(|\vec q\,|)+\frac{|\vec q\,|^2\,{\cal K}^{B'B}(|\vec q\,|)}
{2\left(E_{c}(\vec q\, )+m_c\right)}
\left(\frac{m_c}{E^2_{c}(\vec q\, )}-\frac{1}{m_b}
\right)
\right)\nonumber\\
\frac{-\widehat G_1(|\vec q\,|)+\widehat G_2(|\vec q\,|)+\frac{E_{B'}(-\vec q\, )}{m_{B'}}\,
\widehat G_3(|\vec q\,|)}{E_{B'}(-\vec q\, )+m_{B'}}&=&-\frac{4}{3\sqrt2}\,
\left(\,\frac{{\cal I}^{B'B}(|\vec q\,|)}{E_{c}(\vec q\, )+m_c}
-\frac{{\cal K}^{B'B}(|\vec q\,|)}{2}
\left(\frac{m_c}{E^2_{c}(\vec q\, )}+\frac{1}{m_b}
\right)
\right)\nonumber\\ 
\end{eqnarray} 

\end{itemize}

\section{${\cal I}^{B'B}(|\vec q\,|)$ and ${\cal K}^{B'B}(|\vec q\,|)$
integrals}
\label{app:integrals}
To evaluate  ${\cal I}^{B'B}(|\vec q\,|)$ and ${\cal K}^{B'B}(|\vec q\,|)$
we use  a partial wave expansion of the orbital wave functions:
\begin{eqnarray}
\Psi^{B}_{b\,h_2}(r_1,r_2,r_{12})=\sum_{l=0}^\infty f_l^{B}(r_1,r_2) P_l
(\mu)\nonumber\\
\Psi^{B'}_{c\,h_2}(r_1,r_2,r_{12})=\sum_{l=0}^\infty f_l^{B'}(r_1,r_2) P_l (\mu)
\end{eqnarray}
where $\mu$ is the cosine of
the angle made by $\vec r_1$ and $\vec r_2$ and $P_{l}(\mu)$ is a  Legendre 
polynomial of rank $l$.
The radial functions $f_l^{B}(r_1,r_2),\,f_l^{B'}(r_1,r_2),\,$ are evaluated as
\begin{eqnarray}
f_l^{B}(r_1,r_2)&=&\frac{2l+1}{2}\int_{-1}^{+1}d\mu\
P_l(\mu)\, \Psi^{B}_{b\,h_2}(r_1,r_2,r_{12})\nonumber\\
f_{l}^{B'}(r_1,r_2)&=&\frac{2l+1}{2}\int_{-1}^{+1}d\mu\
P_{l}(\mu)\, \Psi^{B'}_{c\,h_2}(r_1,r_2,r_{12})
\end{eqnarray}
In terms of those we have
\begin{itemize}
\item ${\cal I}^{B'B}(|\vec q\,|)$
\begin{eqnarray}
{\cal I}^{B'B}(|\vec q\,|)&=&16\pi^2\sum_l\sum_{l'}\sum_{l''}
\left(ll'l''\left|\right.000\right)^2
\int_0^\infty d\,r_1\, r_1^2\ j_{l''}(\,\frac{m_{h_2}+m_q}{\overline M'} 
|\vec q\,|r_1)
\int_0^\infty d\,r_2\, r_2^2\ j_{l''}(\,\frac{m_{h_2}}{\overline M'} 
|\vec q\,|r_2)\nonumber\\
&&\hspace{5cm}\times f_{l'}^{B'}(r_1,r_2)\,f_{l}^{B}(r_1,r_2)
\end{eqnarray}
with  $j's$ being spherical Bessel functions.
\item ${\cal K}^{B'B}(|\vec q\,|)$
\begin{eqnarray}
{\cal K}^{B'B}(|\vec q\,|)&=&-\frac{16\pi^2}{\sqrt3\,|\vec q\,|}
\sum_l\sum_{l'}\sum_{l''}\sum_{l'''}\sum_L
(-1)^{(l{''}+l{'''}+1)/2}\sqrt{(2L+1)(2l''+1)(2l'''+1)}\nonumber\\&&
\times\left(ll'l''\left|\right.000\right)\,\left(l''l'''1\left|\right.000\right)
\,\left(l'Ll'''\left|\right.000\right)\,W(l''l'1L:ll''')\nonumber\\&& 
\times\int_0^\infty d\,r_1\, r_1^2\ j_{l'''}(\,\frac{m_{h_2}+m_q}{\overline M'} 
|\vec q\,|r_1)
\int_0^\infty d\,r_2\, r_2^2\ j_{l''}(\,\frac{m_{h_2}}{\overline M'} 
|\vec q\,|r_2)\
 f_{l'}^{B'}(r_1,r_2)\,\Omega_L\,f_{l}^{B}(r_1,r_2)\hspace{1cm}
\end{eqnarray}
with $W(l''l'1L;l,l''')$ being a Racah coefficient and $\Omega_L$ the
differential operator\footnote{Note that the Racah and Clebsch-Gordan
coefficients restrict $L$ to the two possible values  $L=l\pm1$.}
\begin{eqnarray}
\Omega_{L=l+1}&=&-\sqrt{\frac{l+1}{2l+1}}\left(
\frac{\partial}{\partial r_1}-\frac{l}{r_1}\right)\nonumber\\
\Omega_{L=l-1}&=&\ \sqrt{\frac{l}{2l+1}}\left(
\frac{\partial}{\partial r_1}+\frac{l+1}{r_1}\right)
\end{eqnarray}
\end{itemize}
For the actual evaluation we restrict the $l,\,l'$ values to  $l,\,l'=0,\cdots,6$

\end{document}